\documentclass[]{aa}

\usepackage{epsfig}

\usepackage{graphicx}

\usepackage{color}

\usepackage{times}

\usepackage{amsmath}

\usepackage{amssymb}

\usepackage{xspace}

\usepackage{txfonts}

\usepackage{enumitem}

\usepackage{subfig}

\usepackage{hyperref}

\usepackage{float}

\usepackage{natbib}

\def\nar{New. Astro. Review}

\newcommand{\Excursion}  {\mm{{\mathbb E}}}
\newcommand{\Mask}       {\mm{{\mathsf M}}}

\newcommand{\Rspace}     {\mm{{\mathbb R}}} 
\newcommand{\Hspace}     {\mm{{\mathbb H}}} 
\newcommand{\Sspace}     {\mm{{\mathbb S}}}

\newcommand{\relBetti}[1]{\mm{{b}_{#1}}}
 
\newcommand{\relEuler}      {\mm{\sf EC_{\rm rel}}} 
 
\newcommand{\dime}[1]    {\mm{\rm dim\,}{#1}}
\newcommand{\ssx}        {\mm{\sigma}}

\newcommand {\mm}[1] {\ifmmode{#1}\else{\mbox{\(#1\)}}\fi}

\newcommand{\Res}            {\mm{N}}

\newcommand{\Skip}[1]        {}

\newcommand{\npipe}       {{\texttt{PR4 }}}
\newcommand{\ffp}       {{\texttt{PR3 }}}

\newcommand{\tempVarGlobalNpipe}{
	
	\begin{tabular}{|r|r||r|r|r||r|r|r|} \hline
		
		\multicolumn{8}{|c|}{Relative homology -- $\chi^2$ (empirical) -- Gloabal Variance} \\
		\hline
		
		&	& \multicolumn{3}{|c|}{\texttt{North}}  & \multicolumn{3}{|c|}{\texttt{South}} \\ 
		 \hline
		
		Res & \multicolumn{1}{|c||}{FWHM} & \multicolumn{1}{|c}{$\relBetti{0}$} & 
		\multicolumn{1}{c}{$\relBetti{1}$}
		
		& \multicolumn{1}{c|}{$\relEuler$} & \multicolumn{1}{|c}{$\relBetti{0}$} & 
		\multicolumn{1}{c}{$\relBetti{1}$}
		
		& \multicolumn{1}{c|}{$\relEuler$} \\ \hline \hline
		
		\multicolumn{8}{|c|}{threshold = 0.90} \\ \hline
		
		1024	&	10' & 0.4600	&0.1467	&0.2750	& 0.5167&	0.2567	&0.3150		\\ \hline
		
		512	&	20' &   0.1917	&0.1450	&0.1250  & 0.3783	&0.1117	&0.1733	\\ \hline
		
		256	&	40' & 	0.1817	&0.3400	&0.5217& 0.4533	&0.0833	&0.3233 \\ \hline
		
		128	& 80' &	    0.0750	&0.2383	&0.3567&0.1217	& \bf{0.0400}	&0.1367		\\ \hline
		
		64	& 160' &	0.3383	&0.1250	&0.2867	&	0.5150	&0.1583	&0.5817		  \\ \hline
		
		32	&	320' &   \bf{0.0383}&0.2967	&0.2617	&	0.0650	&\bf{0.0017	}&\bf{0.0000}	  \\ 
		\hline
		
		16	&	640' &    0.8800	&0.9617 &	0.9000&	0.3317	& \bf{0.0167}	& \bf{0.0150}				 
		\\ \hline
		
		summary & N/A &0.1667	&0.6383&	0.5283&	 0.0783&	\bf{0.0183}&	\bf{0.0000} \\ 
		\hline

	\end{tabular}
}

\newcommand{\tempVarSepNpipe}{
	
	\begin{tabular}{|r|r||r|r|r||r|r|r|} \hline
		
		\multicolumn{8}{|c|}{Relative homology -- $\chi^2$ (empirical) -- Separate Variance} \\
		\hline
		
		&	& \multicolumn{3}{|c|}{\texttt{North}}  & \multicolumn{3}{|c|}{\texttt{South}} \\ 
		 \hline
		
		Res & \multicolumn{1}{|c||}{FWHM} & \multicolumn{1}{|c}{$\relBetti{0}$} & 
		\multicolumn{1}{c}{$\relBetti{1}$}
		
		& \multicolumn{1}{c|}{$\relEuler$} & \multicolumn{1}{|c}{$\relBetti{0}$} & 
		\multicolumn{1}{c}{$\relBetti{1}$}
		
		& \multicolumn{1}{c|}{$\relEuler$} \\ \hline \hline
		
		\multicolumn{8}{|c|}{threshold = 0.90} \\ \hline
		
		1024	&	10' & 0.4083&	0.0733	&0.1483 & 0.9917&	0.7617	&0.8867			\\ \hline
		
		512	&	20' &0.4633	&0.0850	&0.1650	  & 0.9933	&0.8183	&0.9933		\\ \hline
		
		256	&	40' & 0.0600	&0.3350&	0.4767	&0.3500	&0.8167	&0.7283	 \\ \hline
		
		128	& 80' &	\bf{ 0.0367}	&\bf{0.0367}	& \bf{0.0350}		 &0.6617	&0.9017	&0.9150		  
		\\ \hline
		
		64	& 160' &	\bf{0.0433}	&0.0583	&0.0733	& 0.2833	&0.7217	&0.8500			 \\ \hline
		
		32	&	320' &  \bf{0.0200}	&0.2150	&0.1017		&0.5967	&0.1033	&0.1717		  \\ \hline
		
		16	&	640' &   0.5600	&0.2900	&0.4700	&0.4283	&0.1783	&0.2683		 \\ \hline
		
		summary & N/A & \bf{0.0233}	&0.0533	&\bf{0.0167}	& 0.5767&	0.8767	&0.9250	\\ 
		\hline

	\end{tabular}
}

\newcommand{\tempVarSepffp}{
	
	\begin{tabular}{|r|r||r|r|r||r|r|r|} \hline
		
		\multicolumn{8}{|c|}{Relative homology -- $\chi^2$ (empirical) -- Separate Variance} \\
		\hline
		
		&	& \multicolumn{3}{|c|}{\texttt{North}}  & \multicolumn{3}{|c|}{\texttt{South}} \\ 
		\hline
		
		Res & \multicolumn{1}{|c||}{FWHM} & \multicolumn{1}{|c}{$\relBetti{0}$} & 
		\multicolumn{1}{c}{$\relBetti{1}$}
		
		& \multicolumn{1}{c|}{$\relEuler$} & \multicolumn{1}{|c}{$\relBetti{0}$} & 
		\multicolumn{1}{c}{$\relBetti{1}$}
		
		& \multicolumn{1}{c|}{$\relEuler$} \\ \hline \hline
		
		\multicolumn{8}{|c|}{threshold = 0.90} \\ \hline
		
		1024	&	10' & 0.2400&	0.1100	&0.1333		&0.9667	&0.4567	&0.4667	\\ \hline
		
		512	&	20' &0.4333	&0.2367	&0.2867	 &1.0000	&0.8567	&0.9667\\ \hline
		
		256	&	40' & \bf{0.0367}	&0.3267	&0.2733 & 0.0900	&0.6767	&0.2467	\\ \hline
		
		128	& 80' &		\bf{0.0000}	&\bf{0.0467	}&\bf{0.0133} &0.6533	&0.6233	&0.8467 \\ \hline
		
		64	& 160' &	0.0633	&0.0700	&0.2333		&0.2867	&0.2600	&0.5067 \\ \hline
		
		32	&	320' & 	 0.1567	&0.2467	&0.3433 &0.8533	&0.2467	&0.4600		\\ \hline
		
		16	&	640' &  0.8167	&0.5400	&0.6933&	0.8733	&\bf{0.0233	}&\bf{0.0367}	\\ \hline
		
		summary & N/A &	\bf{ 0.0100}&	0.0867	&\bf{0.0000}	&0.5433	&0.3967&	\bf{0.0133} 
		\\ 
		\hline

	\end{tabular}
}

\newcommand{\tempVarGlobalffp}{
	
	\begin{tabular}{|r|r||r|r|r||r|r|r|} \hline
		
		\multicolumn{8}{|c|}{Relative homology -- $\chi^2$ (empirical) -- Global Variance} \\
		\hline
		
		&	& \multicolumn{3}{|c|}{\texttt{North}}  & \multicolumn{3}{|c|}{\texttt{South}} \\ 
		\hline
		
		Res & \multicolumn{1}{|c||}{FWHM} & \multicolumn{1}{|c}{$\relBetti{0}$} & 
		\multicolumn{1}{c}{$\relBetti{1}$}
		
		& \multicolumn{1}{c|}{$\relEuler$} & \multicolumn{1}{|c}{$\relBetti{0}$} & 
		\multicolumn{1}{c}{$\relBetti{1}$}
		
		& \multicolumn{1}{c|}{$\relEuler$} \\ \hline \hline
		
		\multicolumn{8}{|c|}{threshold = 0.90} \\ \hline
		
		1024	&	10' &0.3400	&0.1400	&0.3100	& 0.3600&	0.1767	&0.2333\\ \hline
		
		512	&	20' & 0.4433	&0.3233	&0.5467	&0.2033	&0.1533	&0.1767\\ \hline
		
		256	&	40' & 0.2067	&0.2167	&0.5433&0.3333	&0.0833	&0.2567	\\ \hline
		
		128	& 80' &	\bf{0.0167}	&0.3033	&0.0867	 &0.2833	&0.0733	&0.1867	\\ \hline
		
		64	& 160' &0.2500	&0.3033 &	0.3000	&0.5733	&0.0933	&0.5100			\\ \hline
		
		32	&	320' & 	 0.1500&	0.5467	&0.5633&0.5067	&	\bf{0.0033}	& 	\bf{0.0033}			\\ 
		\hline
		
		16	&	640' & 0.6400	&0.7167	&0.6500	 & 0.3733	& 	\bf{0.0000}	& 	\bf{0.0000	}	\\ 
		\hline
		
		summary & N/A &	0.2200	&0.4500	&0.4133	& 0.0667&		\bf{0.0000}	& 	\bf{0.0000}	\\ 
		\hline

	\end{tabular}
}

\newcommand{\tempQoneQtwoNpipe}{
	
	\begin{tabular}{|r|r||r|r|r||r|r|r|} \hline
		
		\multicolumn{8}{|c|}{Relative homology -- $\chi^2$ (empirical) -- Separate Variance} \\
		\hline
		
		&	& \multicolumn{3}{|c|}{\texttt{Quad 1}}  & \multicolumn{3}{|c|}{\texttt{Quad 2}} \\ 
		\hline
		
		Res & \multicolumn{1}{|c||}{FWHM} & \multicolumn{1}{|c}{$\relBetti{0}$} & 
		\multicolumn{1}{c}{$\relBetti{1}$}
		
		& \multicolumn{1}{c|}{$\relEuler$} & \multicolumn{1}{|c}{$\relBetti{0}$} & 
		\multicolumn{1}{c}{$\relBetti{1}$}
		
		& \multicolumn{1}{c|}{$\relEuler$} \\ \hline \hline
		
		\multicolumn{8}{|c|}{threshold = 0.90} \\ \hline
		
		512	&	20' &  0.1800	&0.3433	&0.2917	& 0.5617 & \bf{0.0283}	&0.0567		\\ \hline
		
		256	&	40' &  \bf{0.0250}	&0.3650	&0.2233	& 0.5483 &0.5533	&0.7467	\\ \hline
		
		128	& 80' &	 0.4117	    &0.3267	&0.5183	& 0.4350 &0.4767	& 0.5667 \\ \hline
		
		64	& 160' &  \bf{0.0050}	    &\bf{0.0150}	&\bf{0.0033 }& 0.2867 &0.4700	& 0.6450		
		\\ \hline
		
		32	&	320' & 	 0.4283	&0.1517 & 0.4067	&	0.1817	& 0.4083	& 0.6050	\\ \hline
		
		summary & N/A &	 \bf{0.0017}	&\bf{0.0383	}&\bf{0.0100}	&0.2767	&0.3200	&0.2183\\  
		\hline
		
	\end{tabular}
}

\newcommand{\tempQthreeQfourNpipe}{
	
	\begin{tabular}{|r|r||r|r|r||r|r|r|} \hline
		
		\multicolumn{8}{|c|}{Relative homology -- $\chi^2$ (empirical) -- Separate Variance} \\
		\hline
		
		&	& \multicolumn{3}{|c|}{\texttt{Quad 3}}  & \multicolumn{3}{|c|}{\texttt{Quad 4}} \\ 
		\hline
		
		Res & \multicolumn{1}{|c||}{FWHM} & \multicolumn{1}{|c}{$\relBetti{0}$} & 
		\multicolumn{1}{c}{$\relBetti{1}$}
		
		& \multicolumn{1}{c|}{$\relEuler$} & \multicolumn{1}{|c}{$\relBetti{0}$} & 
		\multicolumn{1}{c}{$\relBetti{1}$}
		
		& \multicolumn{1}{c|}{$\relEuler$} \\ \hline \hline
		
		\multicolumn{8}{|c|}{threshold = 0.90} \\ \hline
		
		512	&	20' &  0.9767	&0.6133	 &0.8833   & 0.9283	  &0.6967	&0.9700   \\ \hline
		
		256	&	40' & 0.8100	&0.5233	  &0.7183    &0.2333	&0.7300	 &0.6200	  \\ \hline
		
		128	& 80' &	 0.7983	&0.6400	   &0.8350     &0.7683	   &0.6383	&0.7500   \\ \hline
		
		64	& 160' &	0.5533	&0.4483	  &0.7083	 &0.5933	&0.6100	&0.6267	  \\ \hline
		
		32	&	320' & 	 0.1233	&0.5200	  &0.1433		&0.2100	& 0.3617	&0.3517	\\ \hline
		
		summary & N/A &	0.7433	&0.6200	&0.4683	& 0.4717 &	0.6917	& 0.5533	\\  \hline
		
	\end{tabular}
}

\newcommand{\tempVarSepNpipeMethodB}{
	
	\begin{tabular}{|r|r||r|r|r||r|r|r|} \hline
		
		\multicolumn{8}{|c|}{Relative homology -- $\chi^2$ (empirical) -- Separate Variance} \\
		\hline
		
		&	& \multicolumn{3}{|c|}{\texttt{North}}  & \multicolumn{3}{|c|}{\texttt{South}} \\ 
		\hline
		
		Res & \multicolumn{1}{|c||}{FWHM} & \multicolumn{1}{|c}{$\relBetti{0}$} & 
		\multicolumn{1}{c}{$\relBetti{1}$}
		
		& \multicolumn{1}{c|}{$\relEuler$} & \multicolumn{1}{|c}{$\relBetti{0}$} & 
		\multicolumn{1}{c}{$\relBetti{1}$}
		
		& \multicolumn{1}{c|}{$\relEuler$} \\ \hline \hline
		
		\multicolumn{8}{|c|}{threshold = 0.90} \\ \hline
		
		1024&  10' & 0.3817	 &0.0783	& 0.1500&  0.8850&	0.6650	&0.7083\\ \hline
		
		512	&	20' &  0.6483	&0.1917	& 0.4533 & 0.9900&	0.9033	&0.9967\\ \hline
		
		256	&	40' &  0.0300	&0.4783	  &0.3317 & 0.3683&	0.8417	&0.5933  \\ \hline
		
		128	& 80' &	 0.0467	 &0.0483	&0.0550	&  0.8983	&0.5717	&0.9200 \\ \hline
		
		64	& 160' &	 0.0683	& 0.0400 &	0.1033 &  0.4133	&0.8867	&0.9400	 \\ \hline
		
		32	&	320' & 	 0.0083	 &0.2350	&0.0633	&  0.6650	&0.1367	&0.1900	 \\ \hline
		
		16	&	640' &  0.6150	& 0.3050	&0.5100	&    0.8183	&0.1650	&0.4333\\ \hline
		
		summary & N/A & 0.0117	& 0.1067	& 0.0200 &  0.6933	&0.8267&	0.7350\\   \hline

	\end{tabular}
}

\newcommand{\bettiDiskRing}{
	
	\begin{tabular}{r||r|r|r} \hline
		
		Object & \multicolumn{1}{|c}{$\beta_{0}$} & 
		\multicolumn{1}{|c}{$\beta_{1}$} 
		& \multicolumn{1}{|c}{$\beta_{2}$} \\ \hline
		
		Disk &  1& 0 & 0\\ 
		Ring &  1& 1 & 0\\ 
		Sphere &  1& 0 & 1\\ 
		Torus &  1& 2 & 1\\

		\end{tabular}
		
	}

\newcommand{\skySections}{
	
	\begin{tabular}{r||r|r} \hline
		
		Section & \multicolumn{1}{|c}{polar} & 
		\multicolumn{1}{|c}{azimuthal} \\ \hline
		
		Northern hemisphere &  $0 \le \theta < \pi/2$ & $0 \le \phi < 2\pi$ \\ 
		Southern hemisphere &  $-\pi/2 \le \theta < 0$ & $0 \le \phi < 2\pi$ \\ 
		Quadrant 1 &  $0 \le \theta < \pi/2$ & $0 \le \phi < \pi$ \\ 
		Quadrant 2 &  $0 \le \theta < \pi/2$ & $\pi \le \phi < 2\pi$ \\ 
		Quadrant 3 &  $-\pi/2 \le \theta < 0$ & $0 \le \phi < \pi$ \\ 
		Quadrant 4 &  $-\pi/2 \le \theta < 0$ & $\pi \le \phi < 2\pi$ \\ 
	\end{tabular}
	
}

\begin{document}

\title{Homology reveals significant anisotropy in the cosmic microwave background }

\titlerunning{Evidence for statistical anisotropy in the cosmic microwave background}

\author{Pratyush Pranav\inst{1,2,3},
	\thanks{Email:pratyuze@gmail.com,\\\hspace{1cm}pratyush.pranav@bennett.edu.in}
	\and Thomas Buchert\inst{1}	
}

\authorrunning{Pranav et. al.}

\institute{Univ Lyon, ENS de Lyon, Univ Lyon1, CNRS, Centre de Recherche Astrophysique de 
	Lyon UMR5574, F--69007, Lyon, France	
	\and  School of Artificial Intelligence, Bennett University, Plot no. 8 -12, Greater Noida, UP, India
	\and Ashoka University, National Capital Region P.O., Plot no. 2, Rajiv Gandhi Education City, 
	Rai, Sonipat, Haryana, India	
}

\abstract{
	
	We test the tenet of statistical isotropy of the standard cosmological model via a homology 
	analysis of the cosmic microwave background (CMB) temperature maps in galactic coordinates. 
	The map pixels were
	normalized by subtracting the mean and rescaling by standard deviation, both of which were 
	computed from the relevant unmasked pixels. Examining small sectors of the 
	normalized maps, we find that the results exhibit a dependence 
	on whether we compute the mean 
	and variance locally from the non-masked patch, or from the full masked sky. Assigning 
	local mean and variance for normalization, we find the maximum discrepancy between the data 
	and model in the northern hemisphere, at more than $3.5$ standard deviations (s.d.) 
	for the PR4 dataset at 
	degree scale. For the PR3 dataset, the C-R and SMICA maps display a higher significance than the 
	PR4 
	dataset at $\sim 4$ and $4.1$  s.d., respectively; however, the NILC and 
	SEVEM maps present a
	lower significance at $\sim 3.4$ s.d. The discrepancy is most prominent at scales of 
	roughly a degree, which coincides with the physical scale of the horizon at the epoch of the CMB. The 
	southern 
	hemisphere exhibits a high degree of consistency between the data and the model for both the PR4 
	and PR3 datasets. Assigning the mean and variance of the full masked sky decreases the 
	significance for the northern hemisphere; in particular, the tails.  However, the tails in the 
	southern hemisphere 
	are strongly discrepant at more than $4$ standard deviations at approximately $5$ degrees. 
	The $p$ values obtained from the $\chi^2$-statistic show commensurate 
	significance in both experiments. Examining the quadrants of the sphere, we 
	find the northwest quadrant of the Galactic frame to be the major source of the 
	discrepancy. Prima facie, the results 
	indicate a breakdown of statistical isotropy in the CMB maps; however, more work is needed to 
	ascertain the source of the anomaly. Regardless, these map characteristics may have serious 
	consequences for downstream computations and parameter estimation, and the related problems 
	of 
	Hubble and $\sigma_8$ tension.

}


\maketitle

\section{Introduction}
\label{sec:intro}

The standard Lambda cold dark matter ($\Lambda$CDM) paradigm of cosmology encapsulates 
and 
arises 
from the cosmological principle (CP), which posits that, on large enough scales, the Universe is 
isotropic and homogeneous 
\citep{Milne1936,peebles1970,pee80,bbks,davis1985,hamilton1986,weinberg1987,white1993,liddle2000,harrison2000cosmology,saini2000,Durrer_2015,jones2017precision}.
  Though 
supported by strong 
mathematical, philosophical, and historical foundations,  the veracity of the fundamental tenets of 
CP has not yet been comprehensively and conclusively established, motivating theoretical and 
observational tests \citep{ellis1984,Secrest_2021,Oayda_2023,Dam_2023,Ragavendra_2025}. 
The recent focus of 
cosmology 
toward data 
gathering and 
analysis presents us with an unprecedented opportunity to test the postulates 
of the 
CP, and the ensuing standard model of cosmology. The data gathered 
are from 
both the early and late epochs in the evolutionary timeline of the Universe, and consistently 
present evidence for tensions and anomalies, including the discrepancy in the 
inference of the Hubble parameter and $\sigma_8$ parameter between the data from the early  
and late Universe 
\citep{planck2015cosmoparams,planck2018cosmoparams,Burns_2018,Freedman_2019,
Riess_2021,Balkenhol_2021,Rameez_2021,Abdalla2022,sarkar2022,Perivolaropoulos_2022,aluri2023}.

Cosmic microwave background (CMB) radiation is one of the more important probes of the 
properties of the early Universe. Emitted in the epoch of recombination, when the Universe was 
merely 380,000 years old, the fluctuation characteristics of the CMB trace the fluctuation 
characteristics of the matter distribution and represent an invaluable source 
of information on the properties of the early Universe
\citep{sciama1967,silk1968,bond1984,bond1987,smoot1992,seljak1999a,de_Bernardis_2000,bouchet1999,bouchet2001,jaffe2001,bennett2003,Spergel_2003,
Spergel_2007,planckmain,Durrer_2015,planck2015cosmoparams,jones2017precision,planck2018CMBspectraLikelihood}. They offer the 
largest and oldest canvas on which to test the postulates of CP. Therefore, studying the CMB 
fluctuations is essential for understanding the properties of the stochastic matter field in the 
early Universe.  The two components of the CMB radiation -- temperature and polarization -- 
present independent probes into the properties of the primordial fluctuations 
\citep{seljak1997,zaldarriaga1997,durrer1999,de_Bernardis_2000}.

The  general consensus is 
that the stochastic fluctuation field of the CMB is an instance of 
an isotropic and homogeneous Gaussian random field 
\citep[][see 
also 
\cite{bfs17} and 
references therein for more recent investigations of 
Gaussianity]{harrison1970,guth1980,adler1981,starobinsky1982,guthpi1982,bbks,bouchet2004cmb,komatsu2010}.
However, the 
investigation of CMB data has revealed multiple anomalous features since the  launch of  the Cosmic 
Background 
Explorer (COBE) satellite \citep{mather1991,smoot1991,Wright_1992}. Due to its low 
resolution 
of approximately $7$ degrees, the analysis of 
COBE data first revealed the truly large-scale anomalous lack of correlation in the CMB at 60 
degrees and more \citep{Hinshaw_1996}. The COBE team also pointed out the peculiarity 
of a very low quadrupole moment in the CMB data \citep{mather1991,smoot1992}.
Subsequently, the analysis of data from the  Wilkinson Microwave Anisotropy 
Probe (WMAP) \citep{bennett2003,spergel2007} 
satellite, with a higher resolution, revealed a number of other 
anomalies at smaller scales that have persisted in the CMB measurements by the 
latest Planck 
satellite 
\citep{tegmark2003,eriksen04ng,park2004,planckmain,planckIsotropy2015,planckIsotropy2018}.
 These anomalies, which 
have been detected in both the real and the 
harmonic space, seem to be at odds with the postulates of the standard cosmological model, and 
perhaps with the more fundamental CP itself. 

Representative examples of anomalies in the 
harmonic space consist of the hemispherical power asymmetry (HPA) 
\citep{eriksen2004,Hansen_2004,Hansen_2009,Paci_2010,planckIsotropy2013} or the 
cosmic hemispherical asymmetry (CHA) \citep{Mukherjee_2016}, the alignment of low multipoles 
\citep{tegmark2003,copi2004,schwarz2004,multipoles,cmbanomaliesstarkman}, as well as the 
parity anomaly \citep{Land_2005,Finelli_2012,planckIsotropy2013}.  In particular, the power 
spectrum has been studied at 
large scales, for $\ell = 2,\ldots,40$ \citep{eriksen2004, Mukherjee_2016}, which has later been 
extended to smaller scales, $\ell \sim 600$ \citep{Hansen_2009,Paci_2010,planckIsotropy2013}, 
and the analysis presents the evidence for HPA or CHA at all scales. Important to note is that the 
assumption of cosmological isotropy is challenged in other datasets as well 
\citep{bouchet2001,Colin_2019,Secrest_2021,Secrest_2022,Oayda_2023,Dam_2023}. In 
particular, the 
analysis of galaxy survey datasets also points to a hemispherical asymmetry, as the northern and 
the southern galactic hemispheres appear to have different topo-geometrical characteristics 
\citep{kerscher1997a,kerscher1998,kerscher2001a,appleby2021minkowskiSDSS}.

The anomalies in the 
real space have consisted of the discovery 
of the cold spot \citep{cruzColdSpot}, the anomalous low variance 
\citep{monteserin2008,cruz2011variance}, and the unusual behavior of descriptors 
emerging from topo-geometrical considerations that involve integral-geometric Minkowski 
functionals (MFs) \citep{adler1981,schmalzing1997,sahni1998,pranav2019a}. It is interesting to 
note that while the purely geometric MFs such as the $n$-dimensional volume generally show 
consistency with the model, while the purely topological measures such as the genus and the Euler 
characteristic \citep{gdm86,hamilton1986,weinberg1987,gott1989} hint at deviations from the 
standard model simulations.  Analysis of the CMB maps using MFs was first 
performed on WMAP 
data \citep{schmalzinggorski,park2004,eriksen04ng} and later extended to Planck data 
\citep{planckIsotropy2013,planckIsotropy2015,planckIsotropy2018,pranav2019b,pranav2021loops,pranavAnomalies2021}.
While \cite{park2004} 
performed their analysis on small sub-degree scales, \cite{eriksen04ng} performed a multi-scale 
analysis spanning a range of sub- and super-degree scales. In both cases, there is reported 
asymmetry between the CMB hemispheres. The small-scale analysis in \cite{park2004} 
reports anomalous behavior to the tune of $2\sigma$, while \citep{eriksen04ng} report a more 
than $3\sigma$ deviation in the genus statistic at scales of approximately $5$ degrees for negative 
thresholds. In this context, it is important to note  that the purely geometric entities of 
the MFs, such as the area, contour length, and skeleton length, have consistently shown a 
congruence between the data and the model \citep{planckIsotropy2015,bfs17}, while the 
purely topological entities, such as the genus and the Euler characteristic, have consistently shown 
deviations between 
the data and the model 
\citep{eriksen04ng,park2004,rst,pranav2021loops,pranavAnomalies2021}.

Extending these studies, in this paper, we 
examine specific sectors of  the CMB temperature maps with a view to test the tenet of statistical 
isotropy, via tools that have their basis in 
purely topological notions arising from homology 
\citep{munkres1984,edelsbrunner2010,pranavThesis,pranav2017}. Homology, together with  its 
hierarchical 
extension, persistent homology 
\citep{elz02,edelsbrunner2010,pranavThesis,pranav2017,pranav2019a,pranavReview2021,pranav2021topology2},
 forms the basis 
of the recently emerging field of topological data analysis (TDA) 
\citep{carlsson2009a,wasserman2018,porter2023}. Using these methodologies, which 
form the 
basis for the 
developed computational pipeline tailored to examining the CMB datasets 
\citep{pranavAnomalies2021}, the central contribution of this paper is the uncovering of  a 
number of anomalous signatures, which point to different behavior of galactic 
hemispheres. It is worth noting that evidence for the discovered anomalies has been documented 
in the literature in the past.

The advent of TDA
holds an important place in view of the recent surge in data acquisition in 
cosmology and astronomy, which demands increasingly more sophisticated tools to 
condense meaningful information from these large and growing datasets. 
In view of these observations, the tools and methodologies presented here particularly stand out as  
promising candidates to reveal 
novel features in the big cosmological datasets, including the completed, ongoing, and 
upcoming CMB observations, as is evidenced in this paper, as well as galaxy surveys. Even though a 
recent development, 
TDA has already featured strongly in a wide range of 
astrophysical research, from studies on the large-scale structure of the Universe
\citep{weygaert2011a,pranavde,sousbie1,shivashankar2015,Xu_2019,Kehe2018,codis2018,feldbrugge2019,biagetti2020,kono2020,wilding2020,heydenreich2022,kehe2022,elbers2018,Ouellette_2023} to studies on the stochastic 
properties of astrophysical and cosmological fields in general
\citep{ppc13,rst,makarenko2018,pranav2019a,heydenreich2022,pranav2021topology2}, 
including the CMB fluctuation 
field 
\citep{rst,pranav2019b,pranav2021loops,pranavAnomalies2021}. 

Section~\ref{sec:topology} presents a brief background on the topological concepts. 
Section~\ref{sec:data_and_methods} presents the data and methods, while 
Section~\ref{sec:result_s2_subset} presents the 
main results. We discuss the results and conclude in Section~\ref{sec:discussion}.

\begin{figure}
	
	\subfloat[][]{\includegraphics[width=0.25\textwidth]{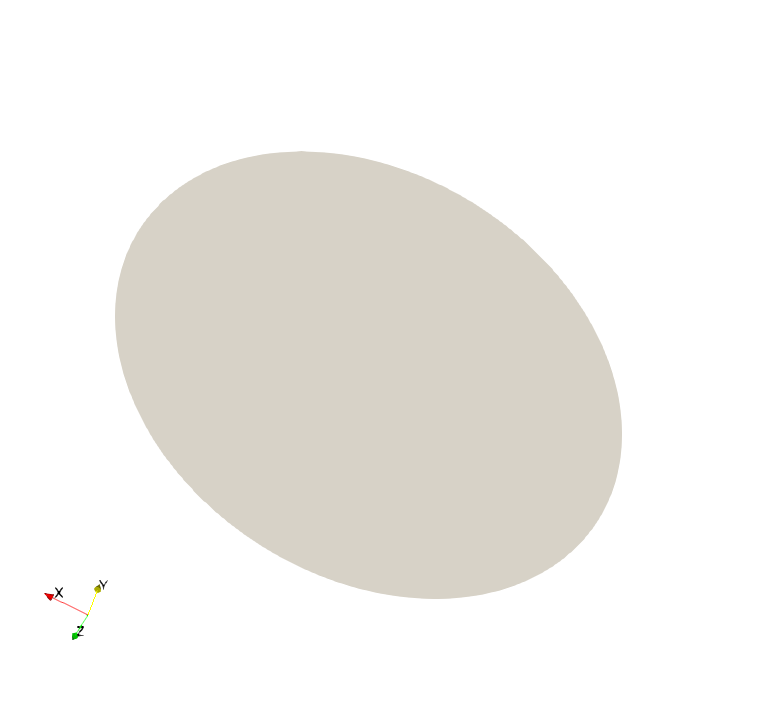}}
	\subfloat[][]{\includegraphics[width=0.25\textwidth]{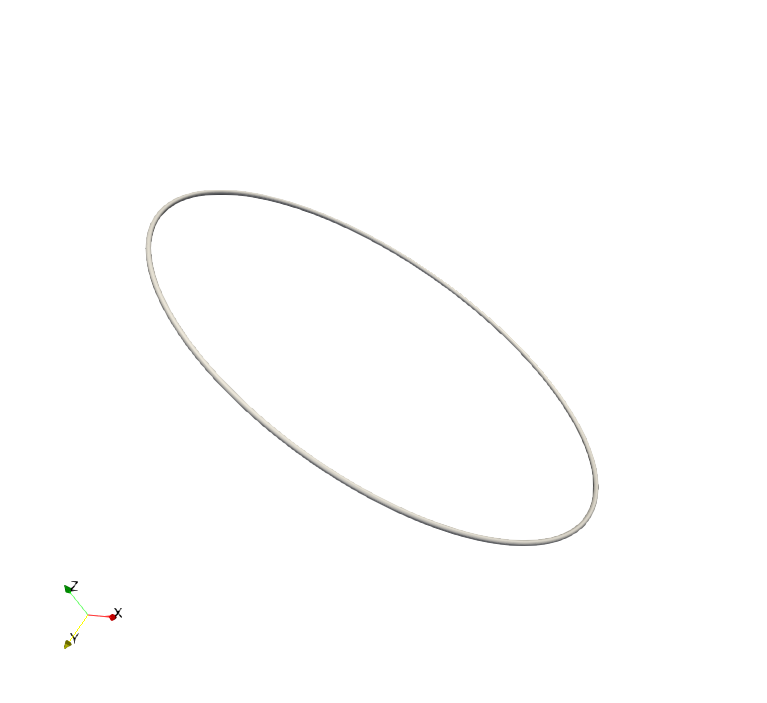}}\\
	\subfloat[][]{\includegraphics[width=0.25\textwidth]{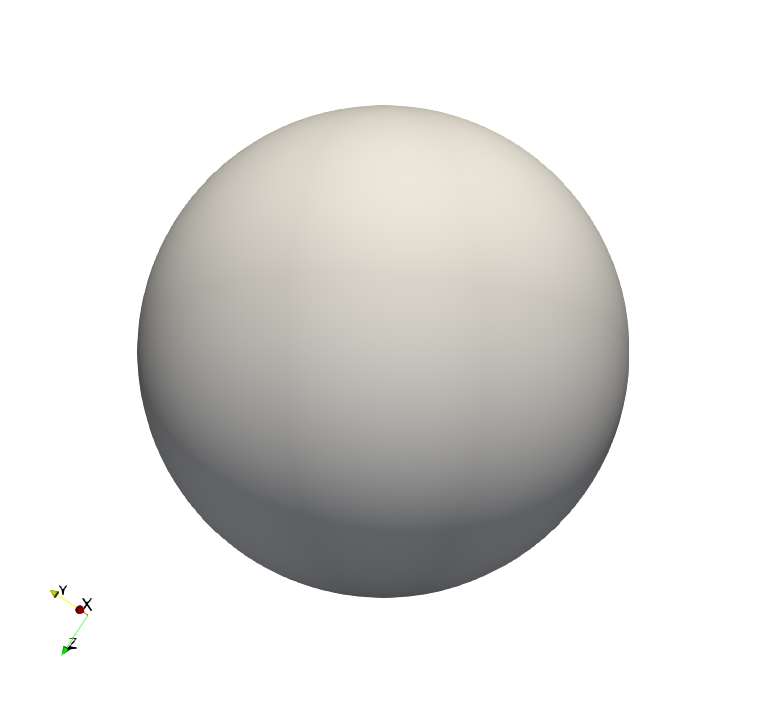}}
	\subfloat[][]{\includegraphics[width=0.25\textwidth]{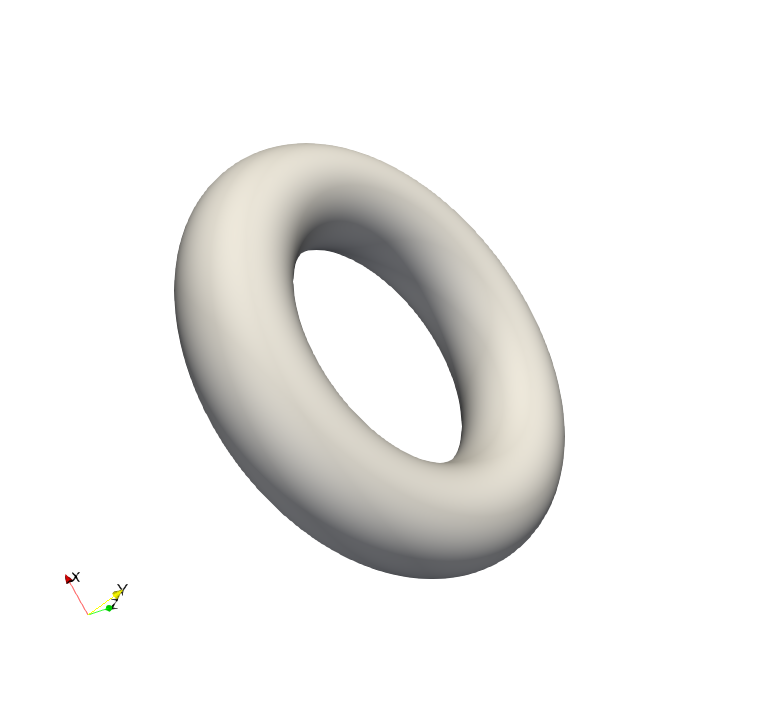}}\\
	\caption{Illustration of objects with different topologies, distinguished by the holes 
		of different dimensions present in them. A disk (panel (a)) is characterized as a single 
		connected object, 
			while a ring  (panel (b)) is characterized as a single connected object that forms the 
			boundary 
			of a hole.  A sphere  (panel (c))  is different from both a disk and a ring, as it is a single 
			connected 
			surface bounding the 3D cavity in the interior, while a torus  (panel (d))  is characterized by 
			a 
			surface in the shape of a hollow tube that surrounds a visible hole, as well as the hole in the 
			interior of the tube body, which also doubles up as a cavity or void.}
	\label{fig:disk_ring}
\end{figure}

\begin{figure*}
	\centering
	\subfloat[][]{\includegraphics[width=0.33\textwidth]{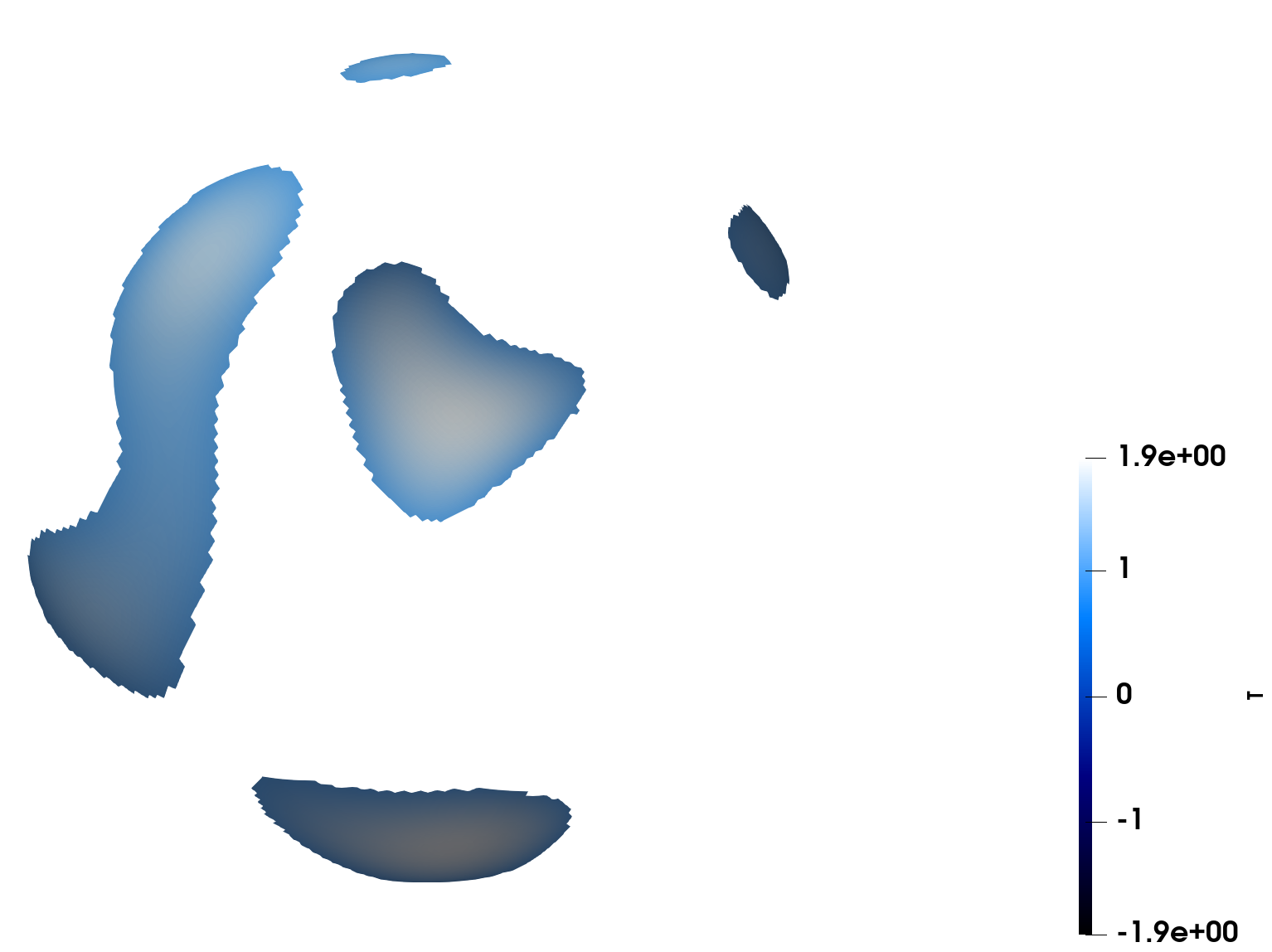}}
	\subfloat[][]{\includegraphics[width=0.33\textwidth]{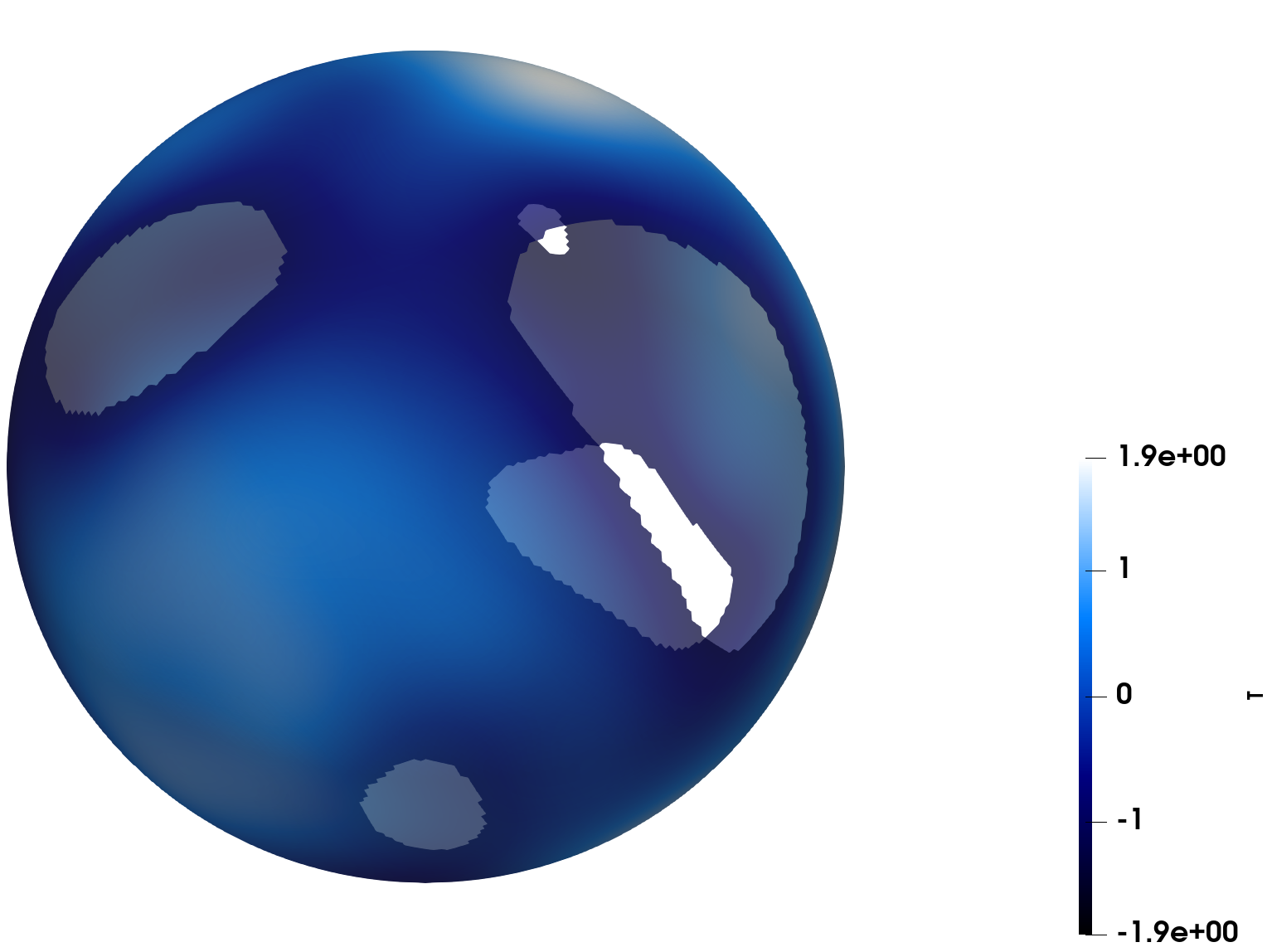}}
	\subfloat[][]{\includegraphics[width=0.33\textwidth]{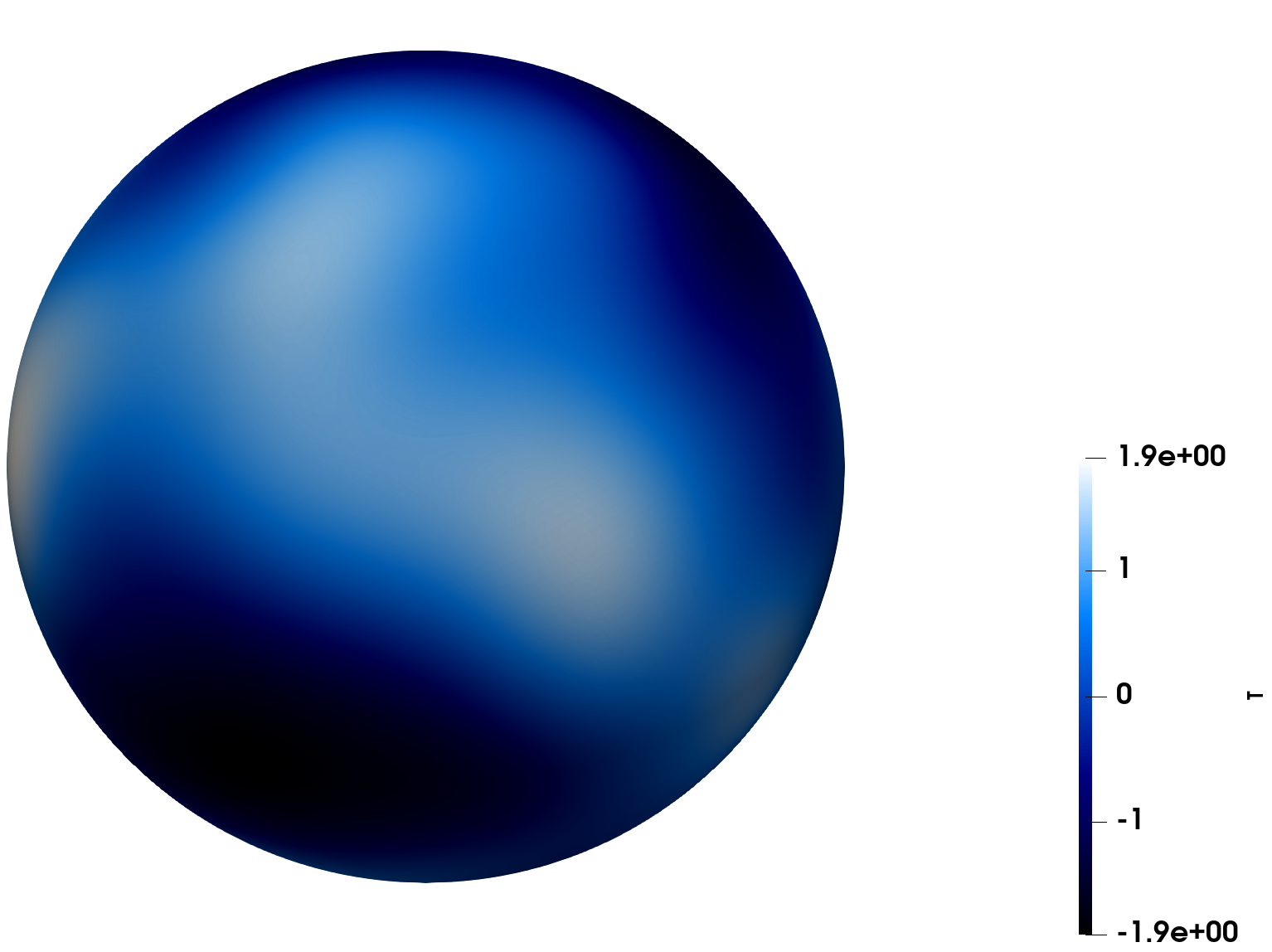}}\\
	\caption{Excursion sets of  the 2-sphere at various thresholds  of the temperature 
		function defined on it. For high positive  thresholds, the excursion set is predominantly 
		composed of isolated objects or components. For low negative thresholds, the 
		excursion set is composed of a single connected component, with additional punctures or loops.
		For low enough thresholds, the excursion set completes to form the full $2$-sphere, composed 
		of a single connected component and a void in the interior.}
	\label{fig:s2_thresholds}
\end{figure*}

\section{Topological background }
\label{sec:topology}

The methods employed in this paper emerge from algebraic and 
computational topology, at the 
level of homology \citep{munkres1984,edelsbrunner2010} and persistent homology 
\citep{elz02,edelsbrunner2010}. In this section, we describe the concepts that are essential for understanding the results presented in this paper in an intuitive fashion, referring the reader to 
\cite{pranav2017} and \cite{edelsbrunner2010} for a more technical presentation.

 These methodologies are complementary in 
nature to the integral-geometric 
Minkowski functionals 
\citep{mecke94,schmalzing1997,schmalzinggorski,eriksen04ng,matsubara2010,ducout2013,bfs17,appleby2018,chingangbam2017,eecestimate},
and together they represent a comprehensive topo-geometrical characterization of fields.

\subsection{Homology}

Homology provides for an exact and unambiguous way of differentiating between  
topological 
spaces by enumerating the holes of different dimensions they contain. As an illustration, we 
examine the different  objects presented in Figure~\ref{fig:disk_ring}, which consist of a 
disk, an annulus, a sphere, and a torus. A disk is a single connected object with no additional 
topological 
structure. It is different from an annulus, because the annulus consists of a single 
connected ring that surrounds a hole. Similarly, the sphere is different from a  disk and an 
annulus, as it consists of a single connected surface that encloses a void. Finally, the torus consists 
of a single connected surface that bounds a void inside the tube, and it also bounds the visible hole. 
In 
addition, the hole inside the tube also serves as a void. Formally, the connected objects, holes, and 
voids, and their generalizations in higher dimensions are associated with the homology group 
$\mathbb{H}_p; p = 0, 1, 2, \ldots, d$, where $d$ is the ambient dimension of the topological space. 
The holes are identified indirectly in homology, by detecting the cycles that bound them. In 
spatial 3D, intuitively, a $0$-cycle is associated with a connected component
\footnote{A 
topological component is different from the components of the measured sky 
signals in the context of observations, in which component separation refers to signal 
processing techniques that extract CMB and foreground from the total measured signals.}, and the 
gap 
between two connected components is the $0$-dimensional hole that they bound. Connected 
components, 
or isolated objects, can also be thought of as a connected cluster of points, and so the 
$0$-dimensional homology cycles are directly associated with clustering properties. A $1$-cycle 
bounds a 
loop or a hole, and $2$-cycle is a connected surface that bounds a 3D void.  The rank of the 
homology group $\mathbb{H}_p$ is the number of independent $p$-dimensional cycles or holes in 
a manifold, and is denoted by the $p$-th Betti number, $\beta_p$ 
\cite{bet71,edelsbrunner2010,pranav2017}. The Betti numbers of the different topological objects 
illustrated in Figure~\ref{fig:disk_ring} are enumerated in Table~\ref{tab:betti_diskRing}.

\begin{table}[htb]
	\centering
	\caption{Betti numbers of the topological objects illustrated in Figure~\ref{fig:disk_ring}.}
	\subfloat{\bettiDiskRing}\\
	\label{tab:betti_diskRing}
\end{table}

We imported the concepts from homology to study the topology of the temperature 
fluctuations of the 
CMB, defined on $\Sspace^2$. In order to do so,  we 
studied the excursion sets of the sphere, defined by the temperature function. For a given 
temperature, $\nu$, the excursion set, $\Excursion(\nu)$, is given by

\begin{equation}
	\centering
	\Excursion(\nu) = \{ x \in \Sspace^2 | f(x) \ge \nu \}.
\end{equation}

On $\Sspace^2$, only the 0- and 1D homology groups are of 
interest. They are associated with isolated components and holes of the excursion sets of 
$\Sspace^2$.  Figure~\ref{fig:s2_thresholds} presents the excursion sets of $\Sspace^2$ at three 
different temperature thresholds.  At high positive thresholds in panel (a), we notice that the excursion 
set is exclusively composed of isolated objects or components. In experimental settings where 
the measurement is performed at discrete locations, these isolated 
objects usually consist of a cluster of points connected to each other, and the number of such 
independent objects is associated with $\Hspace_0$, and represented by the $0$-th Betti number, 
$\beta_0$. At low negative thresholds, we notice that the excursion set is composed of a single 
connected surface, indented by multiple punctures. These punctures are related to 
topological loops or holes associated with the first homology group, $\Hspace_1$, and the number of 
independent such loops counted 
by the first Betti number, $\beta_1$.  At sufficiently low thresholds, the excursion set forms the 
complete sphere, which is a single connected object with no punctures. However, for the complete 
sphere, there is a cavity enclosed by a surface. This cavity is associated with the second homology 
group, $\Hspace_2$, and denoted by the second Betti number, $\beta_2$. 

\subsection{Masks and relative homology}

As was mentioned before, the topological entities on consist 
of components and 
holes, for 
both the excursion sets and the mask. In the presence of the mask, we computed the homology of the 
excursion sets, relative to the mask, and it differs from absolute homology. We did not count a 
component of the 
excursion set completely or partially overlapping with the mask, as it may be 
continuously shrunk into the mask through deformation retraction 
\citep{edelsbrunner2010,pranav2019b}. A hole of the excursion set partially covered by the mask 
was still counted as  a hole, as its boundary completes in the mask. A component of the mask was 
counted as a hole of the excursion set, and a hole of the mask contributed toward a void of the 
excursion set, as in such a case the excursion set is an open disk with its boundary in the mask, and 
therefore is homologous to $\Sspace^2$. A more technical description of relative homology 
can be found in \cite{pranav2019b}.

\subsection{Morse theory and persistent homology}

According to Morse theory \citep{mil63,edelsbrunner2010,pranav2017}, the 
topology of 
excursion sets of a function, $f$, changes only when passing through a critical point of the 
function. A critical point, \textbf{x}, is the location where the gradient of the function vanishes: 
$\nabla f (\bf{x}) = 0$. The critical points are further differentiated by their index, which 
is the number of negative eigenvalues of the Hessian, given by
\begin{equation}
	H_{ij} = \frac{\partial^2 f}{\partial x_i \partial x_j}.
\end{equation}

In 2D, an index-$0$ critical point is a minimum, an index-$1$ critical point is a saddle 
point, and 
an index-$2$ critical point is a maximum. In $d$ dimensions, an index-$k$ critical point either 
gives birth to a $(d-k)$-dimensional topological cycle or  destroys a $k$-dimensional cycle. As an 
example, a saddle point in 2D that has an index of $1$ can have one of the two effects: it either connects 
two separate objects, reducing the number of connected objects by $1$, or it connects  the 
boundary of an already connected object, forming a $1$-dimensional cycle bounding a loop. 
Therefore, each of the topological cycles in the growing excursion set is associated with a pair of 
critical points that are responsible for its birth and death.

The infinite values of thresholds corresponding to the excursion sets of a function 
are 
reduced to a finite number by recognizing that a finite manifold has a countably finite number of 
critical points, and that the topology of the excursion sets remains a constant between two critical 
points. Arranging the finitely many excursion sets corresponding to the  critical points in a 
monotonically decreasing sequence results in a nested sequence of excursion sets known as a 
filtration. This filtration of excursion sets is hierarchical in nature, whereby the excursion set 
at a higher threshold is related to the excursion sets at a lower thresholds through a series of 
inclusion maps, 
which allows for tracking of the birth and death 
of topological cycles through the traversal of excursion 
sets. Persistent homology exploits the hierarchical nature of the filtration and the 
inclusion maps to determine the birth and death thresholds of all the unique topological events that 
occur through the traversal of 
the excursion sets, in terms of the values of the critical points associated with these events. The 
information of persistent homology is represented through 
persistence diagrams or barcodes \cite{elz02,carlsson2005}. We employ the 
former in this paper to 
represent the information of persistent homology. A persistence diagram is a scatter plot in 
$\Rspace^2$, where each of the dots in the diagram is associated with a unique topological feature 
born and destroyed in the filtration of the given function. There is a $p$-dimensional persistence 
diagram for each ambient dimension of the function, where $p=0, 1,\ldots,d$. The Betti numbers for 
a particular threshold, $\nu$, can be extracted from the persistence diagrams. These are precisely 
the topological cycles that are born at or before $\nu$, and that die after $\nu$, or the cycles 
active at the threshold, $\nu$.

\begin{figure}
	\centering
	\subfloat{\includegraphics[width=0.5\textwidth]{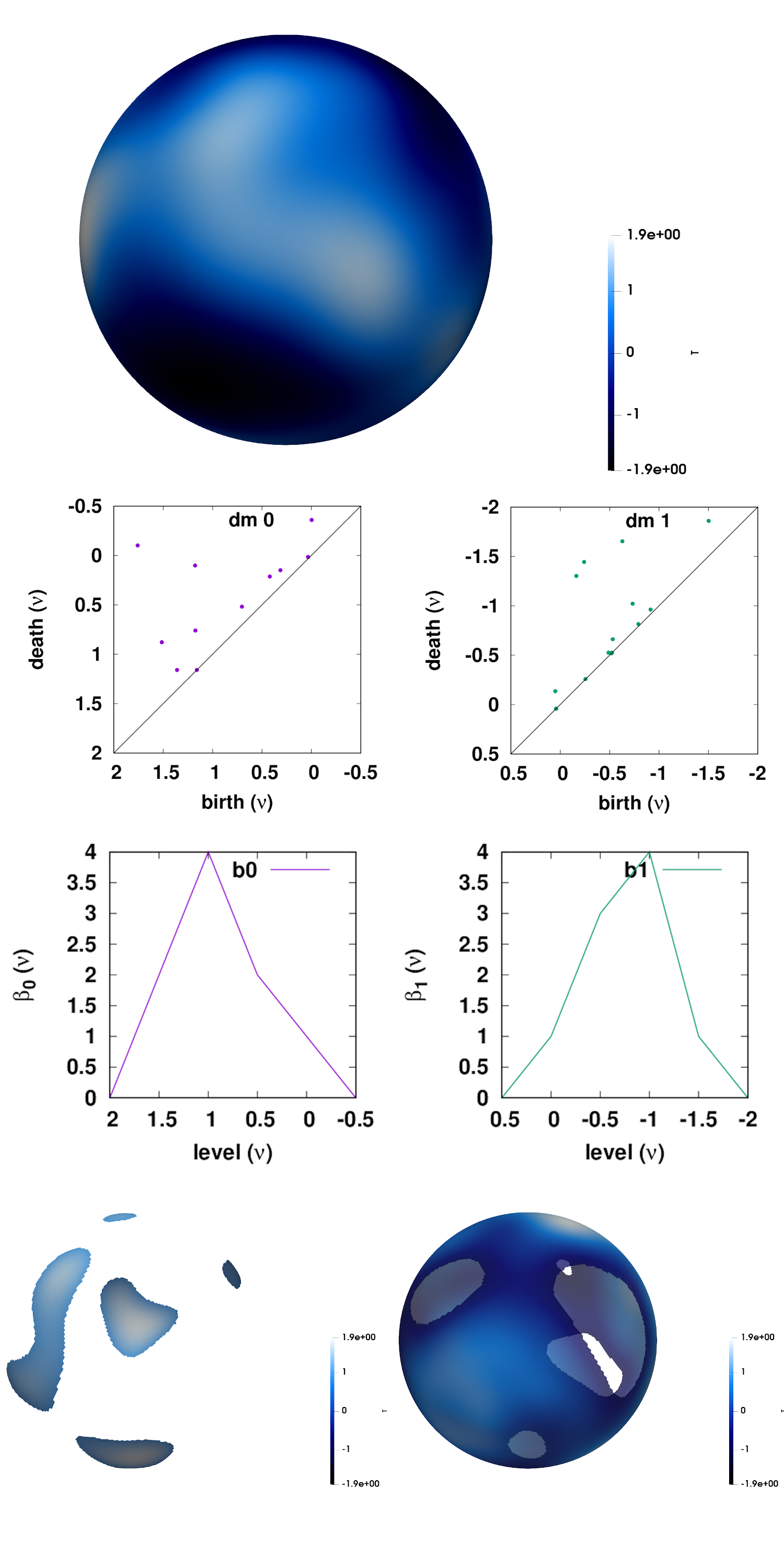}}\\
	\caption{Illustration of the computational aspects of homology and persistent homology. }
	\label{fig:pers_betti_illustration}
\end{figure}

\subsection{Illustration}

Figure~\ref{fig:pers_betti_illustration} presents an illustration of the persistence 
diagrams and 
Betti numbers. The top row of the figure presents the illustration of the field on $\Sspace^2$, 
derived from the CMB temperature fluctuation field smoothed with a Gaussian kernel of a
full width at half maximum (FWHM)$\, = 
20 \deg$. The temperature fluctuation at each pixel was normalized by subtracting the mean and 
rescaling by the standard deviation of all pixels. The two panels of the second row from the top present 
the $0$- and $1$-dimensional persistence diagram, from left to right. In the third row from the top, 
we present the Betti number curves, and in the bottom row we present the excursion sets 
corresponding to $\nu = 1\, \text{and}\, -1$. In the left column, we note the presence of five 
connected components, while in the right we notice a single connected component with five 
punctures.
Correspondingly, in the third row we read from the graphs of Betti numbers, $\beta_0 
(\nu = 1) = 4$, $\beta_1 (\nu = -1) = 4$. In other words, the zeroth and the first Betti numbers have 
values one less than the number of components and the punctures. In the case of components, the 
component born at the highest threshold maps to the full surface at the lowest threshold, which 
belongs to the essential homology of the manifold, and we did not consider this component 
in our computations. In the case of punctures, a sphere with a single puncture on the surface is 
homeomorphic to a disk. Only additional punctures generate holes corresponding to the first 
homology group, and so we 
counted the number of holes as one less than the total number of punctures on the surface of the 
sphere.

\begin{figure}
	\centering
	
	\subfloat[][]{\includegraphics[width=0.25\textwidth]{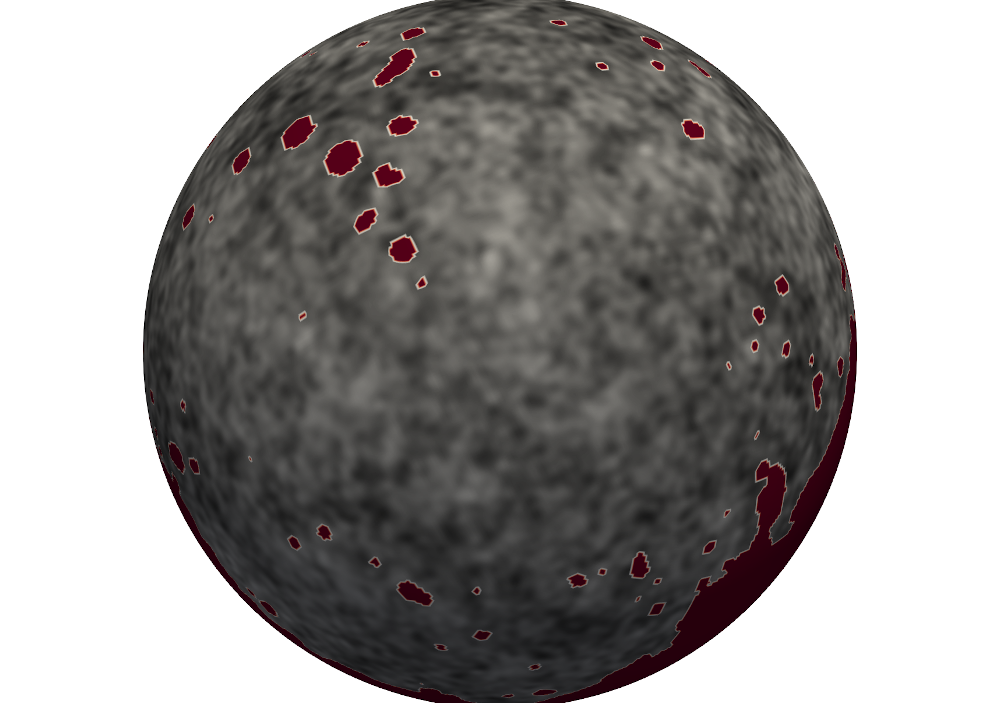}}
	\subfloat[][]{\includegraphics[width=0.25\textwidth]{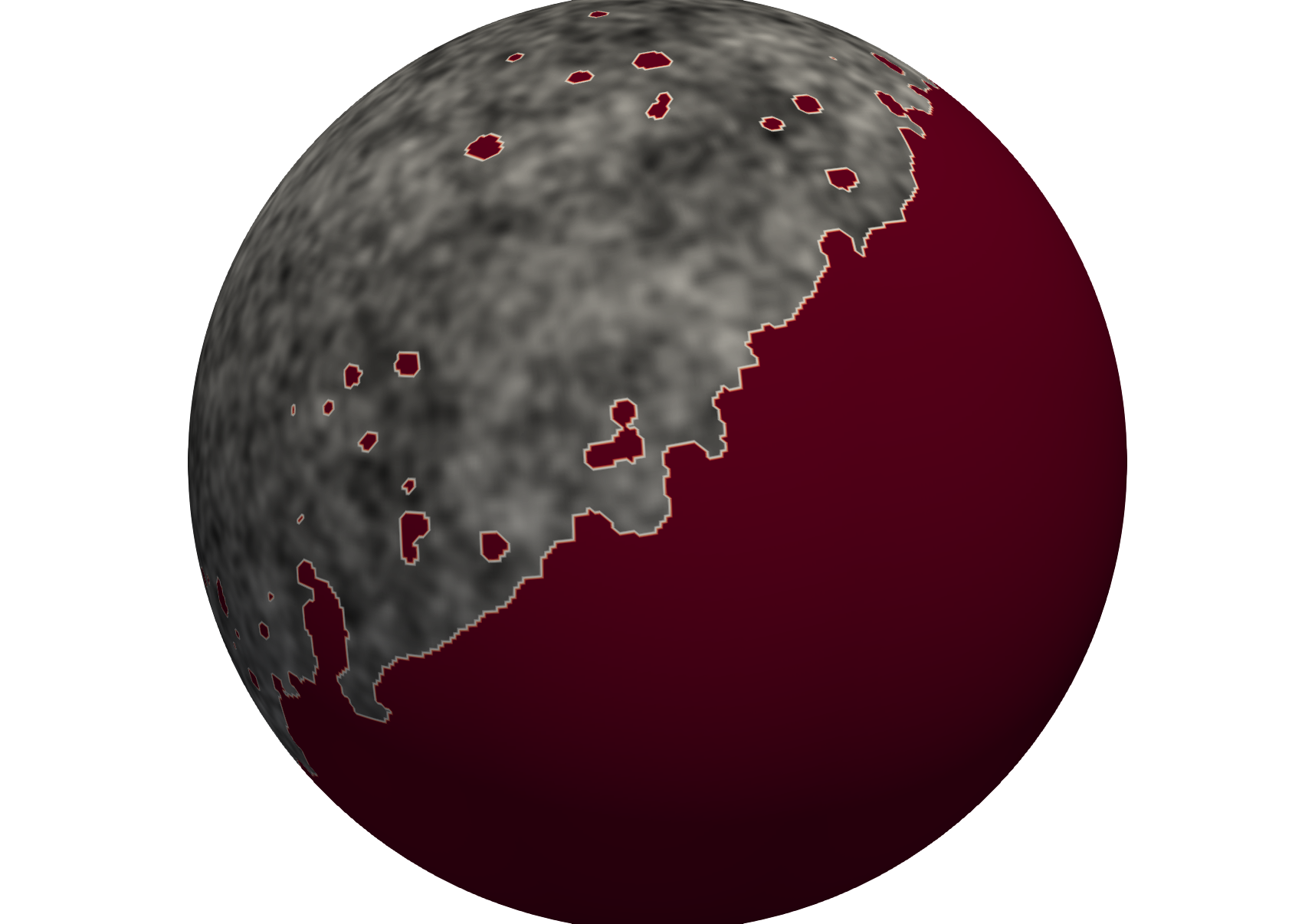}}\\
	
	 \caption{Visualization of the CMB temperature fluctuation field in the northern hemisphere in 
	 	two different views. The masked area covers the whole southern hemisphere and the relevant 
		parts of the northern hemisphere, dictated by the PR3 temperature common mask. The 
		visualization is based on the PR3 observed map, cleaned using the \texttt{SMICA} component 
		separation pipeline, degraded at $\Res = 128$ and smoothed with a Gaussian kernel of $FWHM 
		= 80'$.}
	\label{fig:north_T_vis}
\end{figure}

\section{Data and methods }
\label{sec:data_and_methods}

In this section, we briefly describe the data and methods employed for arriving at the results. All 
the computations were performed using \texttt{TopoS2} \citep{pranavAnomalies2021}, with 
the 
aid of \texttt{HealPix} \citep{healpix1} software for preprocessing. The 
computational pipeline, specifically tailored to the CMB data, but useful 
for the analysis of other scalar functions on $\Sspace^2,$ is a recent development, and a detailed 
account can be found in \cite{pranav2019b} and \cite{pranavAnomalies2021}.

\subsection{Data}
\label{sec:data}

The data that we investigated are the temperature maps from the latest two data releases by the 
Planck team -- the penultimate Planck  Data Release 3 (PR3) \citep{ffp10}, and the fourth and 
final Planck  Data 
Release 4 
(PR4) \citep{npipe}. 
These data releases represent a natural evolution of the Planck data processing pipeline, whereby  
the final data release incorporates the best strategies for both the LFI and HFI instruments, 
commensurate with an overall reduction in noise and systematics \citep{npipe}. The \ffp and the 
\npipe datasets are accompanied by $600$ and $300$ simulations, respectively, which originate 
from the 
standard LCDM paradigm, which posits the CMB field to be an instance of an isotropic and 
homogeneous Gaussian random field. The PR3 dataset consists of observational maps 
obtained via 
four different component separation methods; namely, \texttt{C-R, NILC, SEVEM}, and 
\texttt{SMICA}  (c.f. \cite{planckOverview2018}). We analyzed all four maps in only one 
experiment, to assess the overall trend and congruence of results between the different component 
separation methods. This choice was dictated by significant computational overheads, especially for 
higher resolutions.

\subsection{Computational methods}
\label{sec:methods}

\begin{figure*}
	\centering
		
	\subfloat{\includegraphics[width=\textwidth]{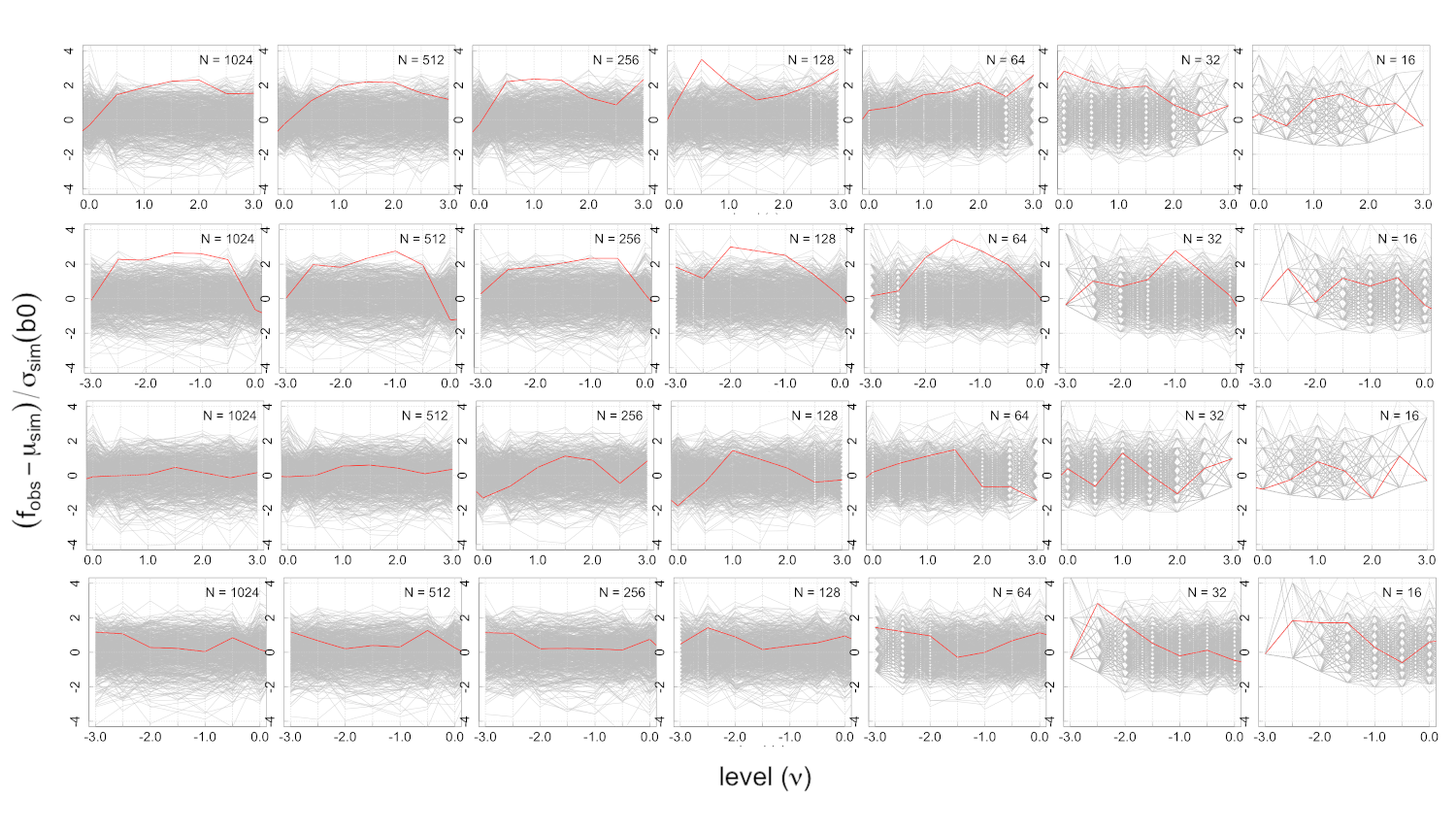}}\\
	
\caption{Graphs of $\relBetti{0}$ and $\relBetti{1}$  for the temperature maps for  the 
	\texttt{NPIPE} dataset  for the northern (top two rows) and southern (bottom 
	two rows) hemispheres. The mean and variance were computed for each hemisphere locally from the 
		unmasked pixels in that hemisphere. The graphs present the normalized differences, and each 
		panel presents the graphs for a range of degradation and smoothing scales. The mask used is 
		the PR3 temperature common mask. }
	\label{fig:npipe_betti_avgSep}
\end{figure*}

\begin{figure*}
	\centering
	\subfloat{\includegraphics[width=\textwidth]{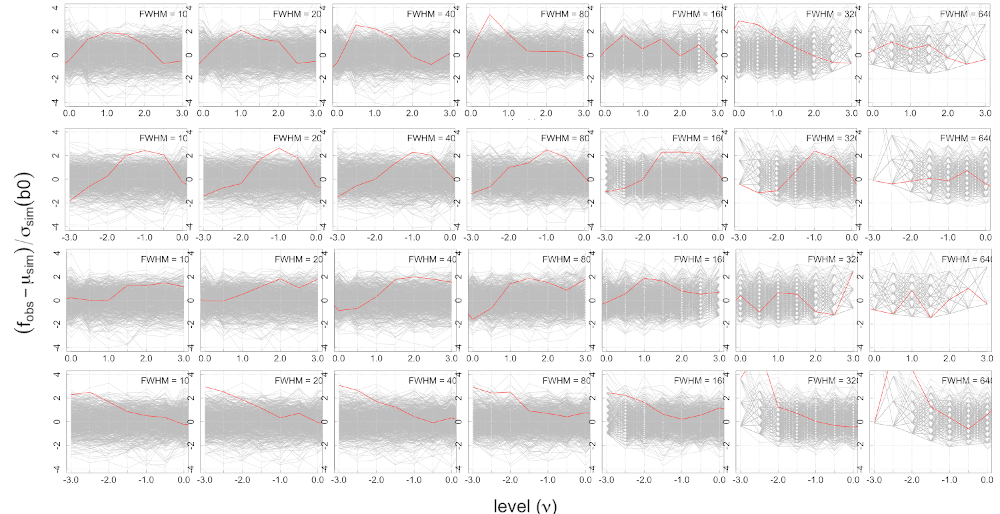}}\\
	\caption{Same as Figure~\ref{fig:npipe_betti_avgSep}; however, in this case, the mean and 
		variance were computed from the full sky from the unmasked pixels.}
	\label{fig:npipe_betti_gAvg}
\end{figure*}

\subsubsection{Masking and preprocessing}

\begin{table}[htb]
	\centering
	\caption{Location of the sections of sky investigated in this paper, in galactic coordinates.}
	\subfloat{\skySections}\\
	\label{tab:skySections}
\end{table}

As the CMB component separation techniques are unreliable in the  regions of strong 
galactic 
emissions, we masked these regions. We 
present our results in 
terms of the homology of the excursion sets relative to the mask, denoting the number of 
components relative to the mask for a given normalized  threshold, $\nu$, as $\relBetti{0} (\nu)$. 
A similar definition was adopted for the number of loops, $\relBetti{1} (\nu)$ 
\citep{pranav2019b,pranavAnomalies2021}. In a continuation of the experiments detailed in 
\cite{pranav2019b}, \cite{pranav2021loops},  and \cite{pranavAnomalies2021}, in this 
paper we 
performed our 
investigations on the 
 hemispheres and quadrants of the CMB sky, defined in galactic coordinates, as is presented in 
 Table~\ref{tab:skySections}.
When analyzing the northern hemisphere, the whole southern hemisphere and the 
relevant parts of northern hemisphere were masked, and vice versa. A similar masking procedure was 
adopted when examining the quadrants. Figure~\ref{fig:north_T_vis} 
presents a visualization of the temperature fluctuation field in the northern hemisphere, smoothed 
with a Gaussian beam profile of $FWHM = 80'$. Figure~\ref{fig:quads} presents a visualization of 
the quadrants of the CMB sky in a Molleweide projection view.

We performed a multi-scale analysis by smoothing the original maps given at $FWHM = 5'$ and $\Res 
= 2048$ to a range of scales defined by a Gaussian beam profile of $FWHM = 10', 20', 40', 80', 160', 
320'$, and $640'$. In order to facilitate faster computations, the maps were also degraded to $\Res = 
1024, 512, 256, 128, 64, 32$, and $16$ in the \texttt{HealPix} format \citep{healpix1}. 
To execute these 
sets of operations, we adopted two different methods, which are briefly described below:
	\paragraph{Method A:} In this procedure, we began by degrading the given maps at 
$N =2048$ to the desired $\Res$. Subsequently, we extracted the spherical harmonic coefficient 
$a_{lm}$ at the degraded resolution, and convolved it with the Gaussian beam at the desired 
FWHM value. Finally, we synthesized the maps from the Gaussian convolved $a_{lm}$ at the final 
output resolution.
	
\paragraph{Method B:} In this procedure, we extracted the spherical harmonic 
coefficients at the input resolution and convolved them with the desired Gaussian beam profile at a 
given FWHM. Subsequently, we synthesized the maps directly at the given output resolution from 
the spherical harmonic coefficients. This is the process that is adopted as a standard, as for 
example 
by the Planck consortium in their analyses of the statistical properties of the CMB 
\citep{planckIsotropy2018}.

We find that the results from the two procedures vary in general, even though they 
exhibit similar characteristics broadly. In particular, method B yields stronger differences 
between the simulations and the observations. In the spirit of conservativeness, we present the 
results from method A in the main section of the paper. We also present a set of results 
from method B in the appendix for comparison.

The mask was subjected to 
an identical degrading and smoothing procedure. This resulted in a nonbinary mask that was 
re-binarized by setting all pixels above and equal to $0.9$ to $1$ and all pixels with smaller values 
to $0$. The mask was applied to the simulations and observations, and they were transformed to 
zero-mean and unit-variance fields by subtracting the mean and rescaling by the standard 
deviation. Denoting $\delta T (\theta,\phi)$ as the fluctuation field at $(\theta,\phi)$ on 
$\Sspace^2$, $\mu_{\delta T}$ as its mean, and $\sigma_{\delta T}$ as the standard deviation, 
computed over the relevant pixels, we examined the properties of the normalized field:

\begin{equation}
	\label{eqn:norm}
	\nu(\theta,\phi)  = \frac{\delta T(\theta,\phi) - \mu_{\delta T}} {\sigma_{\delta T}}.
\end{equation}

In all the experiments, we restricted ourselves to $\nu \in [0:3]$, while 
examining 
$\relBetti{0}$, and $ \nu \in [-3:0]$ while examining $\relBetti{1}$, 
commensurate with the fact 
that 
components are the dominant topological entities at positive thresholds, while loops are 
dominant for the negative thresholds.

\subsubsection{Topology computation}

After the preprocessing steps explained in the previous section, performed with the aid 
of \texttt{HealPix} software 
\citep{healpix1}, 
the data was subjected to the topology computation pipeline, which briefly involves tessellating the 
points on the sphere, computing the upper-star filtration of this tessellation, constructing the 
boundary matrix of the filtration, and reducing the boundary matrix to obtain the $0$- and 
$1$-dimensional persistence 
diagrams. The Betti numbers, relative to the mask, are condensed from the 
persistence diagrams. We describe these steps briefly below, referring the reader to 
\cite{pranav2019b} and \cite{pranavAnomalies2021} for details.

\paragraph{Triangulation} 

We began by projecting the map pixels to $\Sspace^2$, and constructing the triangulation, 
$K$, of 
the set of pixels in 3D. Taking  the convex hull of the triangulation produces a triangulation on 
$\Sspace^2$, which has $V = 12N^2$ vertices, $3V - 6 $ edges, and $2V - 4$ triangles, where $N$ 
is 
the resolution parameter of the map in a \texttt{HealPix} format. This triangulation represents the 
CMB temperature function, $f : \Sspace^2 \to \Rspace$, where the temperature is stored at the 
vertices and all higher-dimensional simplices acquire a piecewise constant interpolation. 

\paragraph{Upper star filtration} 

To construct the upper star filtration, we ordered the simplices of the triangulation such 
that 
$\sigma$ precedes $\tau$ if $f (\sigma) > f(\tau) $ or $f (\sigma) = f(\tau) $ and $dim (\sigma) < 
dim (\tau)$, where $f (\sigma)$ is the minimum temperature value of the vertices of $\sigma$, 
and $dim (\sigma)$ is the dimension of the simplex, $\sigma$. An 
ordering satisfying the above properties constitutes the upper-star filter of $K$ and $f$, and the 
upper-star filtration consists of the prefixes of of the filter, each representing an excursion set of 
$f$. 

\paragraph{Boundary matrix construction, reduction, and persistence computation:}
Given the upper star filtration, and given $\ssx_1, \ssx_2, \ldots, \ssx_n$ as  the sorted 
simplices 
of the upper-star filtration, we defined the  boundary matrix as $\partial [1..n, 1..n]$ as  $\partial [i,j] = 
1$, if $\ssx_i$ is a face of $\ssx_j$ and $\dime{\ssx_i} = \dime{\ssx_j} - 1$, and $\partial [i, j] = 
0$, otherwise. Thereafter, we reduced this ordered boundary matrix to a form commonly 
known as the $lowest (j)$ form. In this matrix reduction method,
a column of the matrix is considered reduced if all its elements in each row are uniformly zero  
throughout. If there are nonzero rows in a column, it is still considered reduced if the lowest row 
with nonzero entry has  only zeros in the same row for all the columns to its left. The row and 
column indices increase from left to right and top to bottom in this convention. In the final reduced 
form, all the columns with nonzero entries have a unique row index for the lowest nonzero entry  
and the row and column indices of these entries determine the birth and death values associated 
with each unique topological feature in the persistence diagram. The 
birth and death coordinates of each dot in the persistence diagram correspond to the values on the 
simplices corresponding to the row and column indices of the $lowest (j) $ (see also 
\cite{pranav2017,pranav2019b}). 

\paragraph{Relative homology computation}
We determined the ranks 
of homology groups relative to the mask from the persistence diagrams by setting the vertices 
belonging to the mask at $+\infty$, through the following set of equations:

\begin{align}
	\relBetti{0}  &= \# \{[b,d) \in Dgm_0(\Excursion \cup \Mask) \mid  +\infty > b \geq \nu > d \} ; 
	\\ \nonumber
	\relBetti{1}  &= \# \{[b,d) \in Dgm_0(\Excursion \cup \Mask) \mid  +\infty = b > d \geq \nu \} \\ 
	\nonumber
	&+ \# \{[b,d) \in Dgm_1(\Excursion \cup \Mask) \mid  +\infty > b \geq \nu > d \} ;\\ \nonumber
	\relBetti{2} &= \# \{[b,d) \in Dgm_1(\Excursion \cup \Mask) \mid  +\infty = b > d \geq \nu \} \\ 
	\nonumber
	&+ \# \{[b,d) \in Dgm_2(\Excursion \cup \Mask) \mid  +\infty > b \geq \nu > d \} .
\end{align}

\section{Results: Topological characteristics of subsets of $\Sspace^2$}
\label{sec:result_s2_subset}
In this section, we present the results of the analysis of various sectors of the CMB sky, with a view 
to testing the statistical isotropy. We begin by presenting the results of examining the hemispheres, 
followed by an examination of the quadrants.

\subsection{Hemispheres}

For the hemispherical analysis, we present the results of two different experiments, which differ 
in 
the regions adopted to compute the mean and variance for normalizing the maps (c.f. 
(\ref{eqn:norm})). In the first 
experiment, we computed the 
mean and variance for each hemisphere  
locally from the unmasked pixels. In the second experiment, we 
computed the mean and 
variance from the unmasked pixels from the hemispheres.  In all the experiments, we present 
our results 
in terms of the graphs of the Betti numbers relative to the mask. The graphs are normalized at each 
threshold to reflect the significance of deviation. At each threshold, we computed the mean, 
$\mu_{\relBetti{i,sim}}$, and the 
standard deviation, $\sigma_{\relBetti{i,sim}}$, of the Betti numbers for $i = 0, 1$, corresponding 
to the components and the holes.  Then, the significance of difference between the 
observation and simulations is given by

\begin{equation}
	\nu_{\relBetti{i}} = \frac{\relBetti{i,obs} -  \mu_{\relBetti{i,sim}}}{\sigma_{\relBetti{i,sim}}},
\end{equation}

\noindent which is the quantity depicted in the graphs.


\begin{table}[htb]
	\caption{Two-tailed $p$ values for relative homology obtained from the 
		empirical Mahalanobis distance or $\chi^2$ test.}
	\small
	\tabcolsep=0.09cm
	\subfloat[][]{\tempVarSepNpipe}\\
	\subfloat[][]{\tempVarGlobalNpipe}\\
	\tablefoot {The table was computed from the sample covariance 
		matrices, 
		for different resolutions and smoothing scales for the \texttt{NPIPE}  dataset. Panel (a) 
		presents 
		the $p$ values for experiments in which the variance was computed for each hemisphere 
		separately, and panel (b) presents results for experiments in which the hemispheres were assigned 
		the variance of the full sky. The last entry is the $p$ value for the summary statistic computed 
		across all resolutions. Marked in boldface are $p$ values of $0.05$ or smaller.} 
	\label{tab:tempNpipe}
\end{table}

\subsubsection{Local normalization}

For these experiments, the temperature at 
each pixel was mean-subtracted, and rescaled by the standard deviation, whereby these quantities 
were computed locally from the unmasked pixels in each hemisphere. The topological 
quantities were 
computed as a function of 
the normalized temperature threshold, $\nu \in [-3:3]$,  in steps of $0.5$. 
Figure~\ref{fig:npipe_betti_avgSep} presents the graphs for the significance of difference between 
the simulations and the observations, for the number of components, $\relBetti{0}$, and the 
number of loops, $\relBetti{1}$, for the \npipe dataset. 
 The red curve represents the observed sky, while the gray 
curves represent the individual simulations treated as observations. The top two rows present the 
graphs for $\relBetti{0}$ and $\relBetti{1}$ for the northern hemisphere, while  the bottom two 
rows present the same for the southern hemisphere. Similar results for the \ffp dataset are 
presented in Figure~\ref{fig:ffp10_betti_avgSep} in the appendix. {In this case, we analyzed all 
four observational maps obtained from the different component separation pipeline.}

We notice a few important things in the graphs: first, that both the \npipe and \ffp datasets present largely identical results; and second, that the significance of deviation in the northern hemisphere is in general flared compared to the southern hemisphere. While the 
southern hemisphere significance is within the $2\sigma$  band in general, the northern 
hemisphere shows a significance of $2\sigma$ or more for most of the scales. The third important 
thing that we notice is the very significant deviation in the number of components between the 
observations and simulations at the threshold $\nu = 0.5$, at $FWHM = 80', \Res = 128$, which is 
approximately at the degree scale.  The significance of the difference for the \npipe 
datasets stands at approximately $3.5\sigma$. For the \ffp dataset, two of the maps, namely 
\texttt{C-R} and \texttt{SMICA}, exhibit a higher significance at approximately $4\sigma$ and 
$4.1\sigma$, respectively, while the two other maps, namely \texttt{NILC} and \texttt{SEVEM}, 
display a $3.4\sigma$ deviation. Figure~\ref{fig:hist_b0_0_5} in the 
appendix presents 
the distribution of Betti numbers at this threshold, which can be approximated as a Gaussian; this justifies ascribing a $\sigma$ significance to the differences.
Figure~\ref{fig:vis_excursion0_5} 
presents the visualization of the structure of the temperature 
field at $\nu = 0.5$ for the northern hemisphere. Clockwise from the top left, we 
present the field for the observed CMB map, as well as three randomly selected simulated samples that are presented in a different color scheme.  
We notice that the observed map exhibits smaller and more fragmented structures compared to the 
simulations. The largest connected structures in the observed and simulated maps are traced by 
connected edges, and it is evident that the simulations display larger connected structures 
compared to the observational map. 

For the loops, the highest deviation 
recorded is at $N = 64, FWHM = 160'$ for the northern hemisphere. The \npipe dataset exhibits a 
$3.4\sigma$ deviation,  while the  \ffp dataset shows a 
$3.2\sigma$ deviation at this scale. 
At $\Res = 128, FWHM = 80'$, the \npipe dataset shows a maximum deviation of
$3\sigma$, while the \ffp dataset exhibits a $2.6\sigma$ deviation.

\subsubsection{Global normalization}

Figure~\ref{fig:npipe_betti_gAvg} presents graphs similar to 
Figure~\ref{fig:npipe_betti_avgSep} for the \npipe dataset, in which the maps are normalized by 
the mean and variance 
computed from unmasked pixels from the full sky.  Similar results for the \ffp dataset 
are presented in Figure~\ref{fig:ffp10_betti_gAvg}. As in the case of local normalization, we notice 
an agreement in the features of the graphs between both datasets. However, we also 
note important differences between the observations and the simulations. For the northern 
hemisphere, the significance of difference is slightly suppressed in general, and there is an 
evident change in the behavior of the tail, where the observation is more consistent with the model. 
However, in this experiment there are significant differences between the observation and the 
model in 
the southern hemisphere. While the number of components, $\relBetti{0}$, is generally within the 
$2\sigma$ band, the number of loops  exhibits strong differences between the data and model in 
the tail. This difference peaks at more than five standard deviations at scales of approximately $5$ 
degrees and more. We note that the distribution of the Betti numbers is manifestly non-Gaussian at 
these scales and thresholds (c.f. \citep{pranavAnomalies2021}).

\subsubsection{Significance of combined thresholds and resolutions}

Table~\ref{tab:tempNpipe} presents the $p$ values from the $\chi^2$ statistics for 
$\relBetti{0}$, $\relBetti{1}$, and 
$\relEuler$ for the \npipe dataset.  $\relEuler$ is the alternating sum of the Betti numbers, and 
indicates the Euler characteristic of the excursion set relative to the mask. The 
$p$ values for the 
\ffp dataset based on \texttt{SMICA} maps are presented in the Appendix  in Table~\ref{tab:ffp}.
The first seven rows in the tables 
present the $p$ values combining different thresholds for a given resolution, accounting for 
multiple testing at different thresholds. For both the datasets, the 
$p$ values are consistent 
with the graphs, broadly indicating similar properties. For experiments in which the variance was 
computed separately for the hemispheres, there is a significant deviation between the data and the model 
approximately at a degree scale, at $FWHM = 80'$, for all the topological quantities. In addition, 
$\relBetti{0}$ exhibits differences for a range of scales at $FWHM = 80', 160', \text{and } 
320'$, with 
$\relBetti{1}$ also displaying mild differences. In 
contrast, the experiments with 
common variance exhibit a departure between the data and the model in the southern hemisphere at 
scales of $5$ degrees and larger. The difference is most evident in the number of loops, which 
also 
affects the $\relEuler$. However, we also note the deviant behavior of $\relBetti{0}$ at $FWHM = 
320'$ in the northern hemisphere.

To account for multiple testing at various resolutions, the last entry in the tables presents the 
summary $p$ values combining all the tested thresholds and resolutions. For the experiments in which 
the variance was computed separately for  different hemispheres, the summary $p$ values present 
significant evidence for nonrandom deviation for all the topological descriptors in the northern 
hemisphere. Similarly, for the experiments with a common variance computed from the full sky, the 
southern hemisphere indicates nonrandom discrepancy between the data and the model for all the 
topological descriptors, which is mild for $\relBetti{0}$ but significant for $\relBetti{1}$ and 
$\relEuler$.

\subsubsection{The effect of normalization}
\label{sec:result_variance}

Normalization is an order-preserving transformation topologically. This 
ensures that there is a bijection 
between
the original and the normalized maps, and specifically a correspondence in the dictionary of critical 
points as well as the order in 
which they appear in the filtration to form and destroy the topological cycles. This entails that, at 
the most,  an 
erroneous estimation of mean and variance would induce an offset between the level sets of the 
compared maps, without affecting the deeper topological structure of the field. 
Combined with the fact that there are stark differences between the data and model irrespective of 
the recipe for normalization, this points to a difference in the topological structure of the 
observed field with respect to the simulations at a level deeper than normalization. 
Appendix~\ref{sec:variance} presents a short account of the properties of the  
distribution of the mean and variance for the full sky and the galactic hemispheres. At the level of 
mean, both the northern and the southern hemispheres are consistent between the data and model, 
which is reflected in the behavior of the mean in the full sky as well. In contrast, the variance of the 
northern hemisphere shows a stark deviation between the observation and simulations, while the 
variance in the southern hemisphere is consistent between the data and the model. This engenders 
a variance in the full sky 
case, which exhibits a mild deviation between the observation and simulations.
Due to the evident discrepancy in variance, it is also prudent to treat both the 
hemispheres as 
arising from different models, and consequently treat the results from this experimental procedure 
in which the mean and variance were computed locally from the hemispheres
as more meaningful. We note that the anomalous behavior of the variance found in our 
analysis is consistent with earlier reports of an anomalous variance \citep{planckIsotropy2013}, which was also 
found in the WMAP data 
\citep{monteserin2008,cruz2011variance}.

\subsection{Quadrants of the sphere}
\label{sec:result_quads}

Following the results of the previous section, in which we detected an anomalous behavior in the 
hemispheres, in this section we compare the behavior of the observations and simulations in the 
different quadrants of the sphere for the \npipe dataset,  illustrated in Figure~\ref{fig:quads}, with 
a view to determine 
the 
zone of discrepancy more accurately. In these experiments, we computed the mean and the variance 
locally from the quadrants for map normalization. We restricted ourselves to the analysis of scales 
represented by  $FWHM = 20,40,80,160, \& 320$. Our choice of scales was determined by the fact 
that at larger scales statistics on smaller sections of the sphere may not be reliable due to the low 
numbers involved, while the 
smaller scales have significant computational overheads.

Figure~\ref{fig:betti_4quads_npipe_avgSep} presents the graphs for the Betti numbers, while 
Table~\ref{tab:tempNpipe_quads} presents the $p$ values computed from the empirical $\chi^2$ 
test for the different quadrants, based on $600$ simulations.  Examining the graphs, the first 
quadrant stands out due to the 
maximum deviation for $\relBetti{1}$  at $FWHM = 160'$, with a significance of more than 
$3.8\sigma$. The $p$ values 
presented in Table~\ref{tab:tempNpipe_quads} corroborate the fact that  the northern hemisphere 
exhibits anomalous behavior with respect to the simulations, where the source of the discrepancy is 
clearly associated with the first quadrant. The summary $p$ values combining all thresholds and 
resolutions indicate a strong nonrandom deviation between data and model in the first quadrant, 
which is strongest for $\relBetti{0}$. Interesting to note is that the deviations between the data 
and the model for the quadrants have shifted to a slightly higher resolution of $FWHM = 160'$, where  $\relBetti{0}$ also exhibits the strongest deviations. In comparison, the quadrants of the southern 
hemisphere exhibit no difference 
with respect to the model, which is consistent with the observation that the southern Galactic hemisphere is 
congruent with the standard model, when the maps are normalized by local mean and variance. 

\begin{table}
	\caption{Two-tailed $p$ values for relative homology obtained from the 
		empirical Mahalanobis distance or $\chi^2$ test}
	\small
	\tabcolsep=0.09cm
	\subfloat[][]{\tempQoneQtwoNpipe}\\
	\subfloat[][]{\tempQthreeQfourNpipe}\\
	\tablefoot{The values were computed from the sample covariance 
		matrices, 
		for different resolutions and smoothing scales  for the maps in different quadrants of the sphere 
		for the \texttt{NPIPE}  dataset. Panel (a) 
		presents the $p$ values for experiments in which the variance was computed for each hemisphere 
		separately, and panel (b) presents results for experiments in which the hemispheres were assigned 
		the variance of the full sky. The last entry is the $p$ value for the summary statistic computed 
		across all resolutions. Marked in boldface are $p$ values of $0.05$ or smaller.} 
	\label{tab:tempNpipe_quads}
\end{table}

\subsection{Comparison with earlier results in literature}

\paragraph{Local hemisphere variance}

In the experiments in which the variance was computed locally from the hemispheres, the most striking 
feature 
is the  significant deviation at a degree scale in the northern hemisphere. We have noticed hints 
of this degree-scale deviation in the full sky analysis reported in 
\cite{pranavAnomalies2021}, in which we reported a $2.96\sigma$ deviation in the \ffp dataset at 
$FWHM = 80'$. The \npipe dataset at this scale exhibits a $2.2\sigma$ deviation. Due to a weak 
display of anomaly in the \npipe dataset, we rejected it in view of the larger scales exhibiting 
stronger anomalies statistically. 

We find a similar phenomenon in the investigations on the WMAP data reported in the literature. 
\cite{park2004} and \cite{eriksen04ng} pioneered the investigation of genus 
statistics of WMAP CMB data. While \cite{park2004} stuck to 
small sub-degree scales, \cite{eriksen04ng} performed a multi-scale  
topo-geometrical investigation spanning a range from sub-degree small scales to super-degree large 
scales.  Due to 
its multi-scale analysis, as well as the fact that it was performed on WMAP data and represents 
independent evidence, we find \cite{eriksen04ng} an excellent 
source for comparison with our results. Figure 4 of \cite{eriksen04ng} presents the Minkowski 
functional curves for the WMAP data smoothed at $FWHM = 1.28\deg$, where the genus in the 
positive threshold range deviates from the simulations at more than $2\sigma$. This is more 
evident 
in the Figure 5 of \cite{eriksen04ng}, in which the MFs and the skeleton length are presented for a 
range of smoothing scales. We notice a  discrepancy in the genus between observations 
and simulations  at more than $2\sigma$ for positive thresholds for a range of scales, most 
prominently around $1.28\deg$ 
and $1.70\deg$. The genus for the negative thresholds shows no 
anomaly at these scales. This also results in the suppression of the signal from the positive 
threshold anomaly in 
the $\chi^2$ statistic for the genus. Simultaneously, for the larger scales of approximately $5$ 
degrees, the genus at negative 
thresholds deviates by more 
than $3\sigma$, and therefore, like in our case in \cite{pranavAnomalies2021},  
\cite{eriksen04ng} also do not deem the degree scale deviation to be significant in view of the 
anomaly at larger scales. In the same paper, Figure 10 shows that the asymmetry parameter 
between 
negative and positive thresholds for the genus is weakly correlated. This is experimental evidence 
in support of the theoretical fact  that the different Betti numbers, which dominate the genus at 
different thresholds, are independent, which further motivates the case for examining the Betti 
numbers separately, in addition to their linear combination, reflected in the genus. Our experiments 
further support this observation, as  the 
Betti numbers of the hemispheres reveal a difference in their topological properties, which is the 
source of weak deviation observed in the full sky analysis for the Betti numbers, as well as the 
associated genus statistics, in both the Planck and WMAP data.   

\paragraph{Global variance}

Assigning the global variance to the hemispheres gives rise to discrepant behavior between 
observations and simulations in the southern hemisphere, specifically for the loops at scales of 
roughly $5$ degrees and larger. The observed discrepancy stands at approximately $5$ standard 
deviations. We have also observed hints of this phenomenon in the full sky analysis, in which 
we report a difference of $3.9$ standard deviation in the number of loops between the observations 
and 
simulations at this scale \citep{pranavAnomalies2021}. The source of this deviation is linked to the 
strongly deviant behavior of loops in the southern hemisphere. Moreover, this phenomenon is also
observed in the WMAP data, in which the genus at negative thresholds exhibits large deviations 
from the simulations at these scales \citep{eriksen04ng}.

\section{Discussions and conclusion}
\label{sec:discussion}

In this paper, we have presented a multi-scale analysis of the topological properties of the CMB 
temperature maps in small sectors of the sky, including hemispheres and quadrants, with the aim of 
investigating the veracity of the postulate of statistical  isotropy. 
This is a continuation of the experiments performed in 
\cite{pranav2019b}, \cite{pranav2021loops}, and \cite{pranavAnomalies2021}, in which we report 
on the full sky 
properties of the temperature fluctuation maps. 
We have employed tools emanating from homology and its hierarchical extension, persistent 
homology, 
which form the the foundations of computational topology  
\citep{elz02,edelsbrunner2010,pranav2017}. 

We have found various anomalous signatures in the topology of the 
temperature fluctuations  in the normalized maps in small sections of the sky defined in the 
Galactic coordinates. In the case of hemispheres, the characteristics of the discovered 
anomalous signatures depend on the section of the sky adopted 
to compute the mean and variance for normalizing the maps. For experiments in which the 
hemispheres were assigned a local mean and variance for normalization, we find that the northern 
hemisphere 
shows significant deviations between the data and the model, most prominently  at scales of 
roughly a degree, in which case the coincidence between the scale of the anomaly and the horizon at 
the epoch of CMB is worth noting. The anomalous signatures in this case are more prominent for 
topological components; however, the topological holes also present anomalous behavior. In 
contrast, 
the southern hemisphere displays a remarkable consistency with the standard model simulations. 
For the experiments in which the hemispheres were assigned a global mean and variance computed 
from the 
masked full sky, the southern hemisphere exhibits strongly anomalous behavior with respect to the 
standard model simulations at scales of roughly 5 degrees and more. 

Noting that the variance of the 
northern hemisphere is starkly different, while the variance of the southern 
hemisphere is strongly consistent with the model, it may be prudent to treat the hemispheres as 
arising from different models at the level of normalization. Consequently, the results from the 
experiments in which the variance was computed locally from the hemispheres may be a more accurate 
reflection of reality than the experiments in which the hemispheres were assigned a global variance for 
normalization. In view of this, the degree-scale anomalies in the behavior of the topological 
components in the northern hemisphere may be a stronger and fairer indicator of the ground truth,  
compared to the larger-scale anomalies in the topological loops in the southern hemisphere that 
could be artifacts of data treatment. Taking hints from the experiments on 
the 
hemispheres, we further tested the quadrants of the sphere assigning local variance for 
normalization. In this case, we find that the first quadrant displays significantly anomalous 
behavior with respect to the simulations. 

Despite possible
offsets due to an erroneous estimation of 
mean and variance,  considering the fact that normalization is an order-preserving 
transformation, and that the anomalies persist irrespective of the recipe adopted for normalization, 
this points to deeper differences in the stochastic structure of the observed and simulated CMB 
fields, at a level beyond normalization and offset effects. The fact that the deviations persist in the 
$\chi^2$ tests, which take into consideration all the thresholds and resolutions, is another 
compelling argument against the deviations being engendered by offset effects. However, to avoid 
and mitigate any such 
effects, in future research we shall present a comparison of topology directly in the space of 
persistence diagrams, which encode consolidated information about all level sets. Preliminary 
inroads in this direction have been been made in terms of modeling 
persistence diagrams directly, with an analysis of the Galactic  hemispheres in 
\cite{rst}.  

An 
agnostic interpretation of these data characteristics points to a departure from statistical 
isotropy in the CMB maps; however, further work is required to convincingly ascribe the source of 
the anomalies to a genuinely cosmological effect, a foreground effect
\citep{bouchet1999}, or merely 
systematic effects. Regarding noise and systematics, there is an important point to 
consider from first principles. The \npipe dataset purportedly has a lower level of systematics and 
noise compared to the \ffp dataset. If this is the ground truth, and if the anomalous signals that we have 
discovered are real, their significance should increase from \ffp to the \npipe datasets. We notice this 
trend for the \texttt{NILC} and \texttt{SEVEM} maps, but the opposite trend for \texttt{C-R} and 
\texttt{SMICA} maps, which makes any assessment about the origin of the signals inconclusive. 
It is also important to note that  the slightly differing results for the different maps in \ffp indicate 
the sensitivity of our methodology to the details of the component separation pipeline, and can be 
considered as a framework for benchmarking. 

In 
the context of foregrounds \citep{bouchet1999}, the 
recently discovered foreground effect by 
\cite{hansen2023possible} also deserves a specific mention. They find that there is an effect of 
deepening of the CMB temperature profile extending to a few degrees around nearby large spiral 
galaxies. They posit that this foreground effect may provide a possible common explanation 
for a number of anomalies of different kinds in the CMB. In the context of the topological anomalies, 
such an effect may generate spurious holes in addition to deepening their temperature profiles. 
However, we find it difficult to conclude that the aforementioned foreground can account for the 
novel anomalous signatures presented in this paper, for a few reasons. First, the evidence 
presented 
in this paper points to strongly anomalous signatures in the hot spot regions also, which cannot 
be explained by the deepening of temperature profiles, or the creation of spurious cold spots. Second,  the 
fact that an anomalous number of 
cold spots appear in the southern hemisphere when assigning a global variance for normalization, 
and disappear when 
normalizing the hemispheres by local variance, points to the fact that the anomalous behavior of 
the cold spots in the southern hemisphere may in fact be misleading, and engendered as an artifact 
of data processing. The third 
compelling reason is that we find the deviations to be the most significant in the first quad in 
galactic coordinates, and the foreground map presented in \cite{hansen2023possible} lacks 
significant foreground contamination in this region. Regardless 
of the finer details, a crucial point to note from the evidence presented in this paper is that the 
disagreement  
between the data and the model may engender spurious results for all subsequent downstream 
calculations, such as cosmological parameter estimation
\citep{Fosalba_2021,Yeung_2022}, and may have consequences for the Hubble and $\sigma_8$ 
tension through 
the 
related misestimation of cosmological parameters.

\begin{acknowledgements}

We thank the anonymous referee for the numerous incisive comments and suggestions 
that have improved the manuscript significantly. We are indebted to  Herbert Edelsbrunner, Rien 
van de Weygaert, Armin Schwartzman, and  Robert Adler for discussions 
and comments that have 
helped shape the draft. We are grateful to Hans Kristian Eriksen, Mohammed Rameez and Geraint 
F. Lewis for important 
comments on the 
draft. 
PP would also like to acknowledge the crucial interactions with Reijo Keskitalo and Julian Borill, 
their consistent help with questions, and incisive comments on the draft. 

PP is partially supported by the Simons-Ashoka fellowship (grant no: 993444).
This work is part of  a project that has received funding from the European 
Research Council (ERC) under the European Union's Horizon 2020 research and innovation 
programme (grant agreement ERC advanced grant 740021 - ARTHUS, PI: TB). A part of this work 
was 
supported by the Physics of Living Matter Group at the University of Luxembourg, and by the 
Luxembourg National Research Fund's ATTRACT Investigator Grant no. 
A17/MS/11572821/MBRACE, and CORE Grant no. C19/MS/13719464/TOPOFLUME. We gratefully 
acknowledge the support of PSMN (P\^ole Scientifique de Mod\'elisation 
Num\'erique) of the ENS de Lyon, and the Department of Energy’s National Energy Research 
Scientific Computing Center (NERSC) at Lawrence Berkeley National Laboratory, operated under 
Contract No. DE-AC02-05CH11231, for the use of computing resources.

\end{acknowledgements}

\begin{appendix}

\onecolumn

\section{Additional figures }
\label{sec:additionalfigures}
\begin{figure*}[h!]
	\centering
	\subfloat[][]{\includegraphics[width=0.6\textwidth]{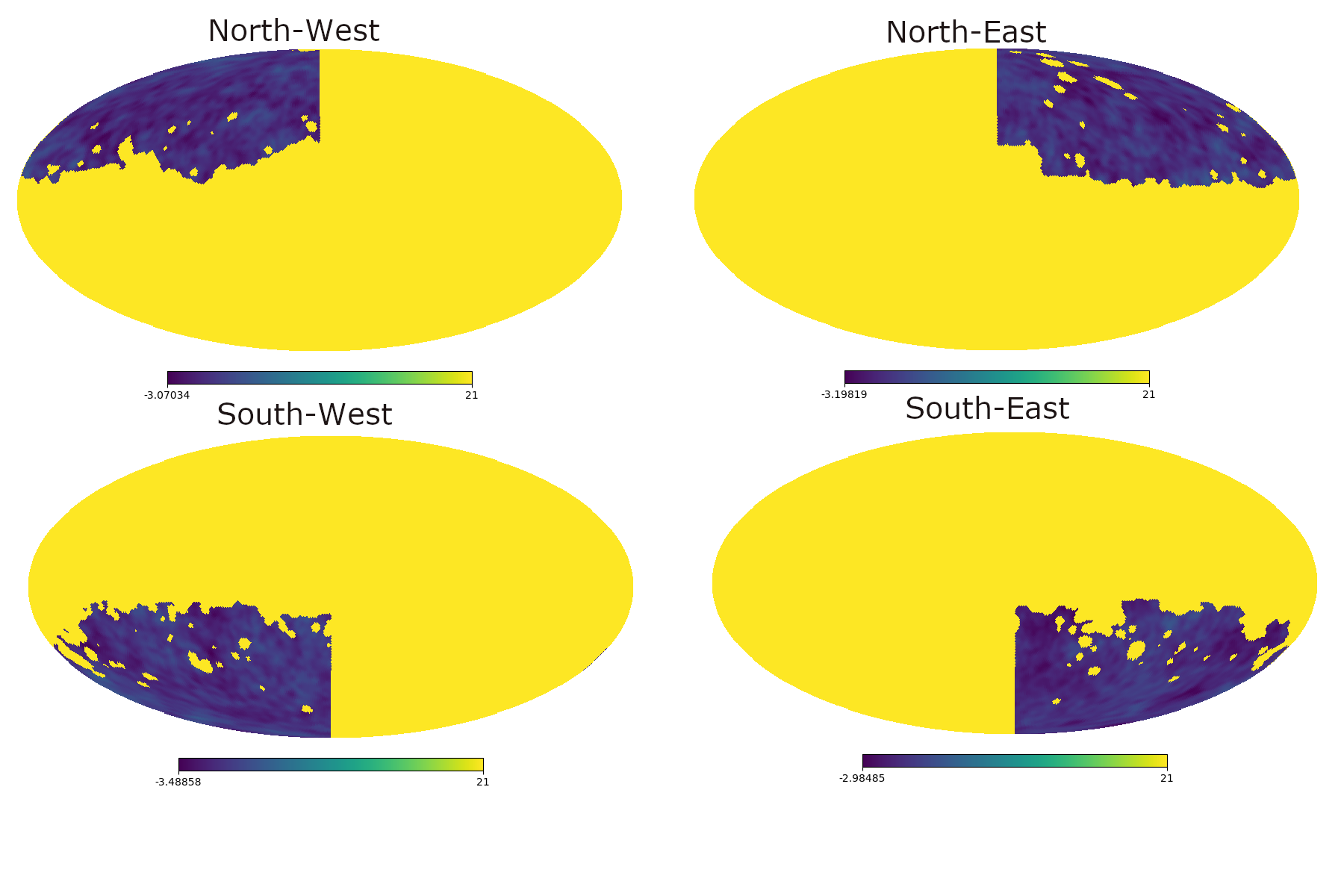}}\\
		\caption{An illustration of the analyzed map surface in the  different quadrants of the sphere. 
		The first, second, third and fourth quadrants are presented from left to right, and 
		this convention is followed for quoting the results.}
	\label{fig:quads}
\end{figure*}

\begin{figure*}[h!]
	\centering
	
	\subfloat{\includegraphics[width=0.24\textwidth]{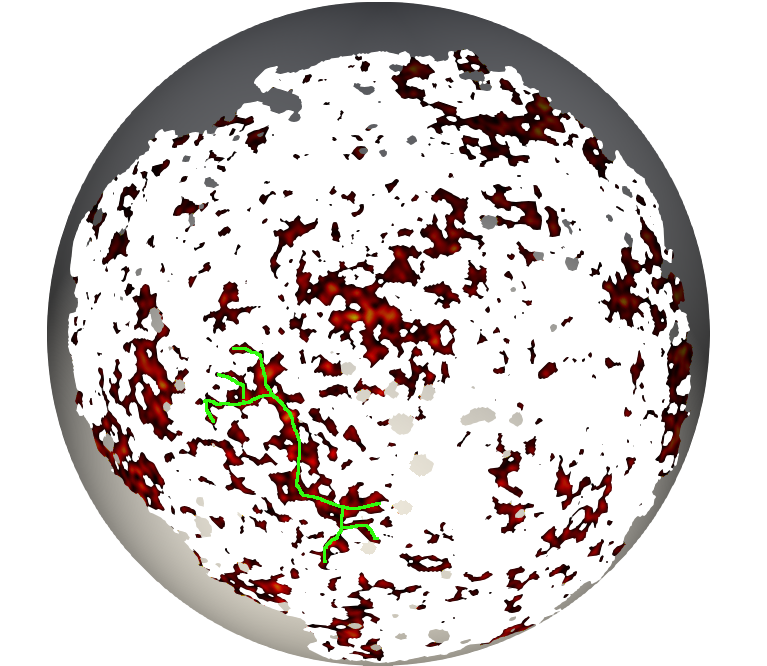}}
	\subfloat{\includegraphics[width=0.24\textwidth]{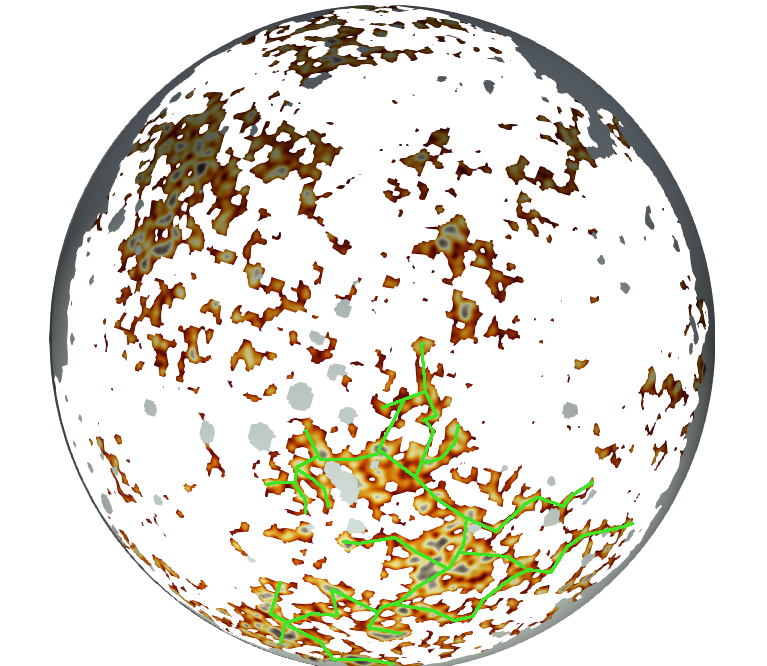}}
	\subfloat{\includegraphics[width=0.24\textwidth]{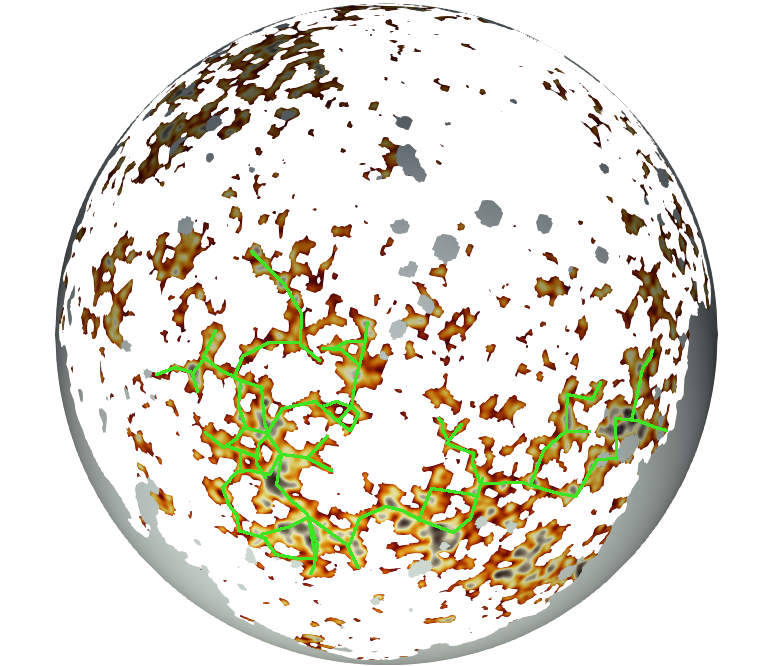}}
	\subfloat{\includegraphics[width=0.24\textwidth]{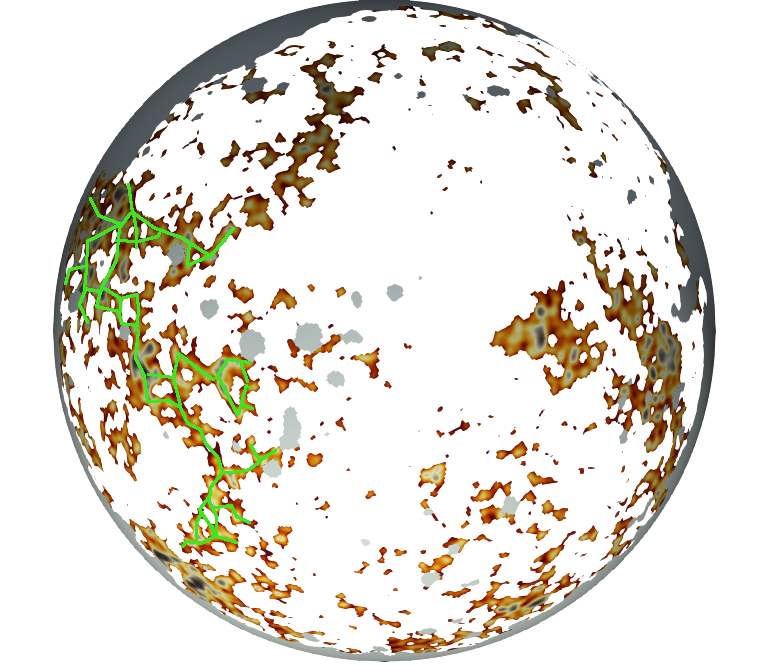}}
	
	\caption{A visualization of the structure of the superlevel set of the temperature field 
		at the  threshold $\nu = 0.5$ for the northern hemisphere. This is the threshold at which we 
		detect 
		a statistically significant deviation between the observation and simulations in the number of 
		isolated components.  The top-left panel presents the visualization of the 
		observed CMB map from PR3 data release. The rest of the panels present the visualization for 
		the randomly selected 
		simulation sample from the \texttt{FFP10} simulation set, numbering 42, and its two higher 
		multiples 
		84 and 126. All the maps are smoothed with a Gaussian beam profile of 
		$FWHM = 80'$. At the level of visual examination, we find that the observational map is 
		composed of larger number of smaller structures compared to the simulations, which display 
		the evidence of larger structures. The largest connected structure in the observational and 
		simulated maps are traced by green connected segments for reference. It is evident that the 
		simulations are characterized by larger such structures compared to the observational map.}
	\label{fig:vis_excursion0_5}
\end{figure*}

\begin{figure*}[h!]
	\centering
	
	\subfloat[][]{\includegraphics[width=0.6\textwidth]{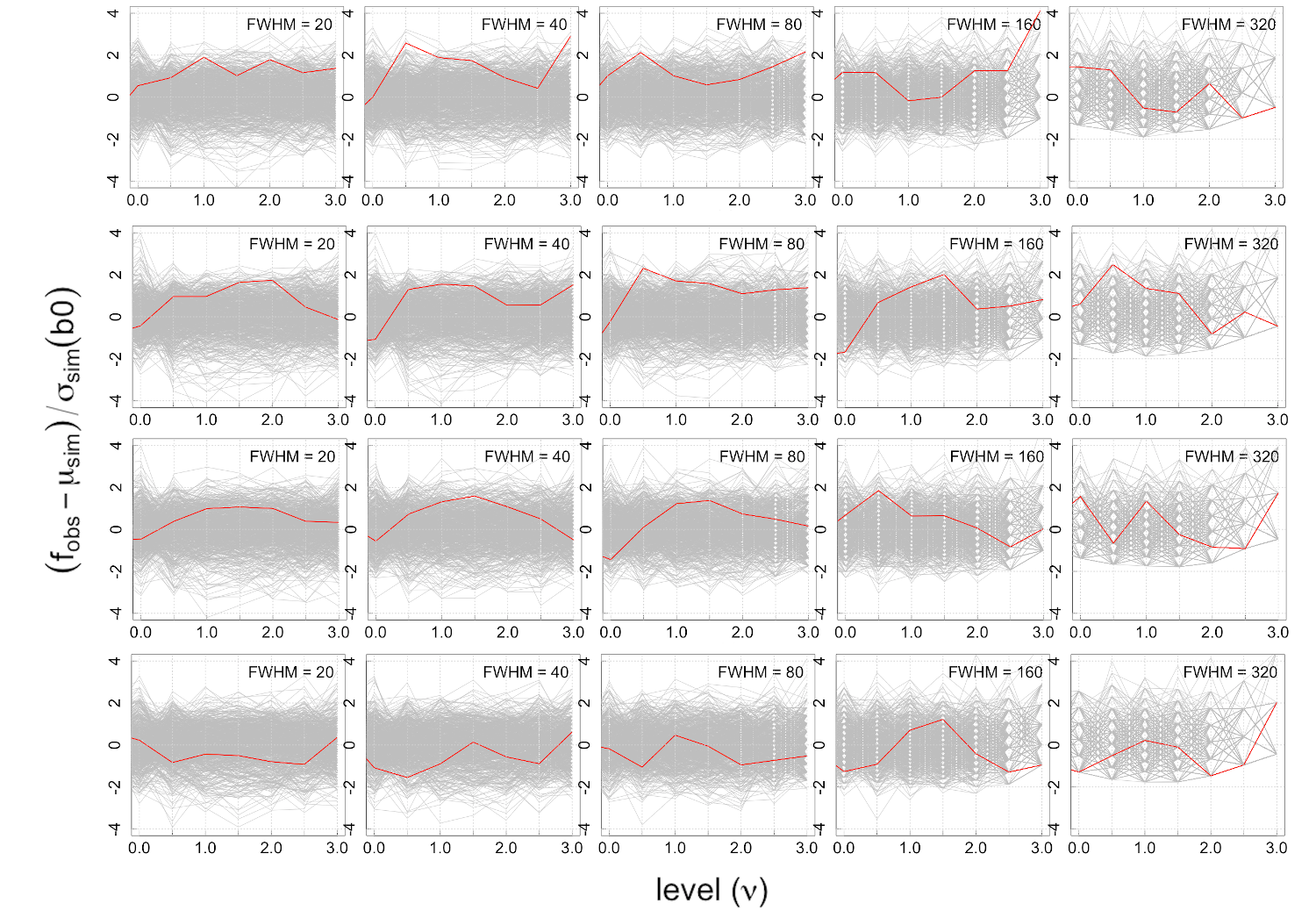}}\\
	
	\subfloat[][]{\includegraphics[width=0.6\textwidth]{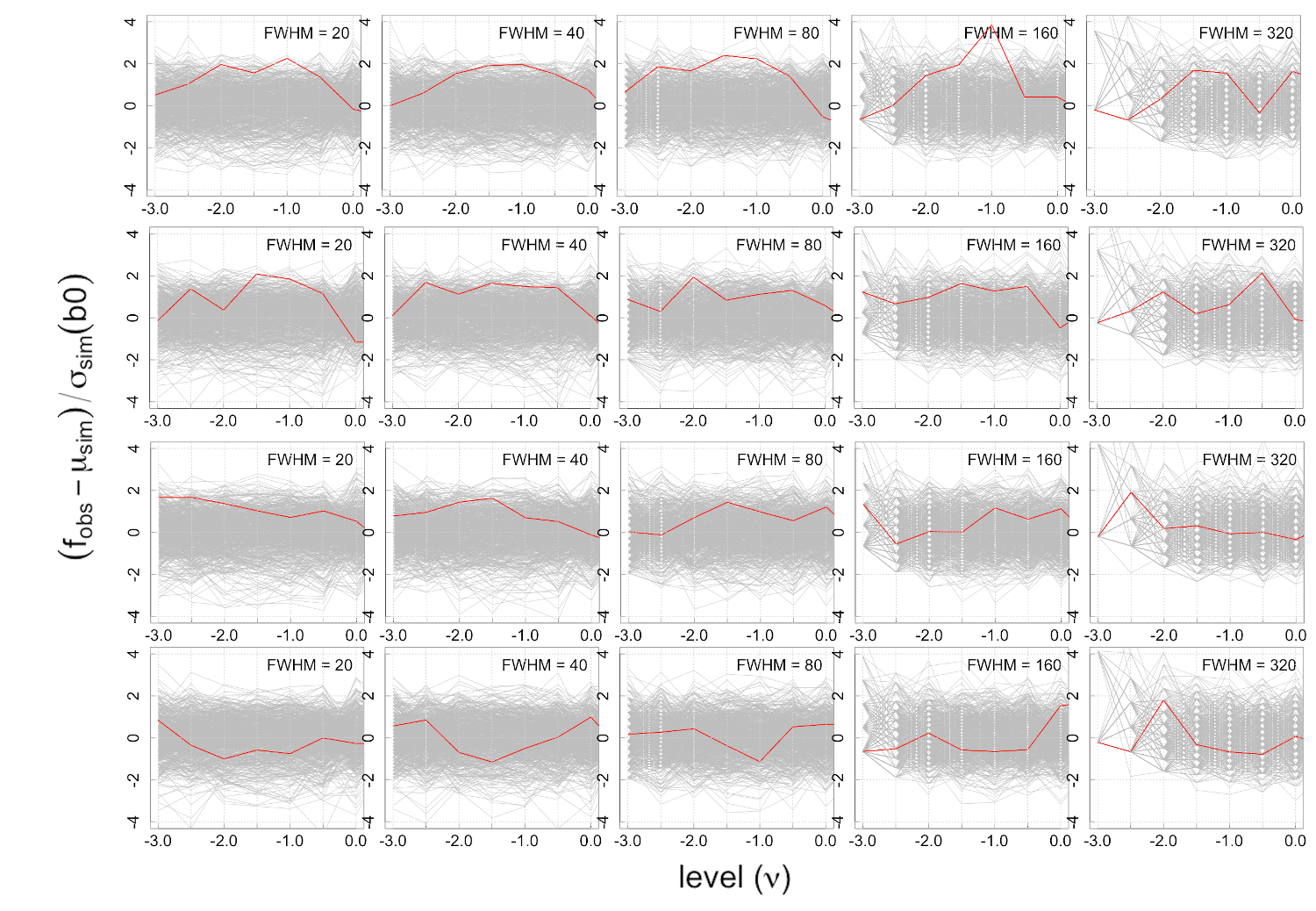}}\
	
	%
	%
	%
	%
	
	  \caption{Graphs of $\relBetti{0}$ and $\relBetti{1}$ for the different quadrants of the sphere 
	  	for a range of smoothing scales with Gaussian $FWHM = 20', 40', 80', 160', 320'$. Panel (a) 
		presents the graphs for $\relBetti{0}$ while panel (b) presents the 
		graphs for $\relBetti{1}$. The values for the different quadrants are presented from top to 
		bottom, while the scale increases from left to right.}
	\label{fig:betti_4quads_npipe_avgSep}
\end{figure*}

\clearpage

\section{Distribution characteristics of mean, variance and Betti numbers}
\label{sec:variance}

In the results presented in the previous sections, we have noticed features in 
the topological characteristics that exhibit  weak to strong dependence on the 
recipe for computing mean and variance for normalizing the maps. Specifically, computing the 
mean and variance locally 
from the hemispheres  points to a difference between the data and model in the northern 
hemisphere, as opposed to the southern hemisphere that shows remarkable consistency with the 
model. In contrast, computing 
the variance from the full sky results in a deviation between the data and model for both the 
hemispheres for some scales. 

\begin{figure*}[h!]
	\centering
	
	\subfloat[][]{\includegraphics[width=0.25\textwidth]{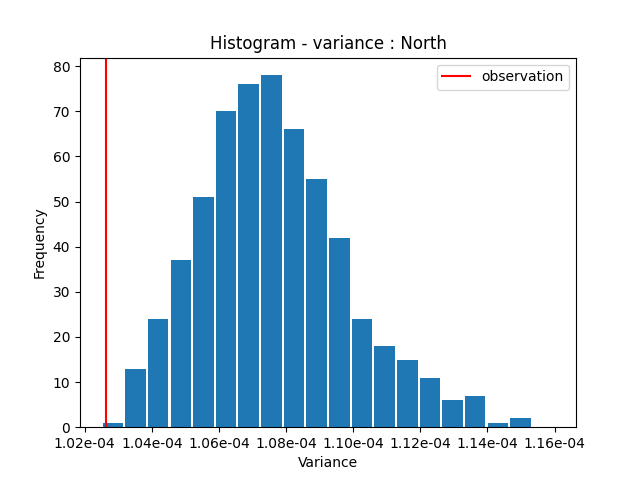}}
	\subfloat[][]{\includegraphics[width=0.25\textwidth]{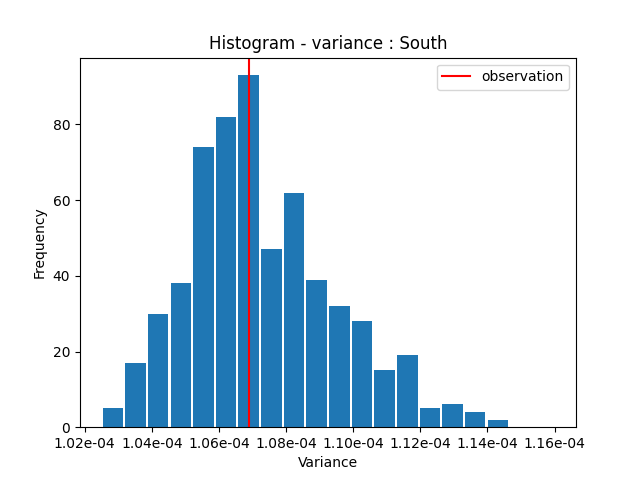}}
	\subfloat[][]{\includegraphics[width=0.25\textwidth]{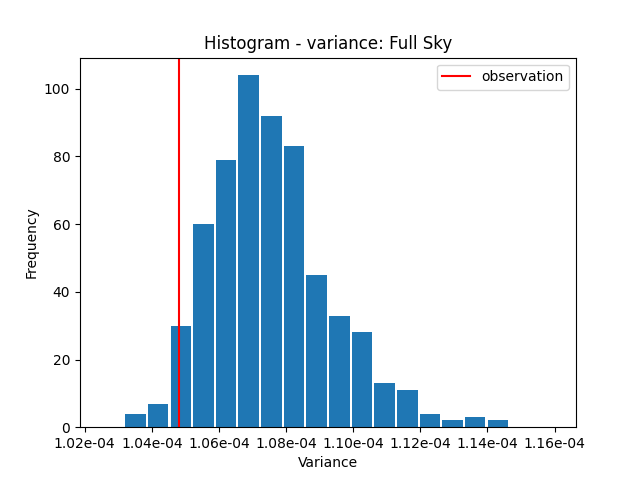}}\\
	\subfloat[][]{\includegraphics[width=0.25\textwidth]{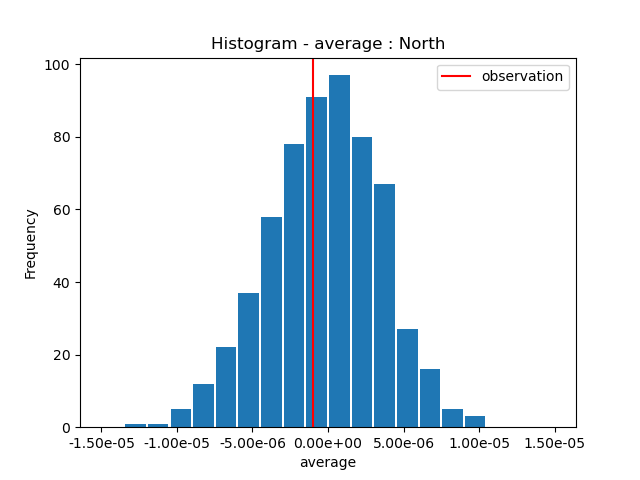}}
	\subfloat[][]{\includegraphics[width=0.25\textwidth]{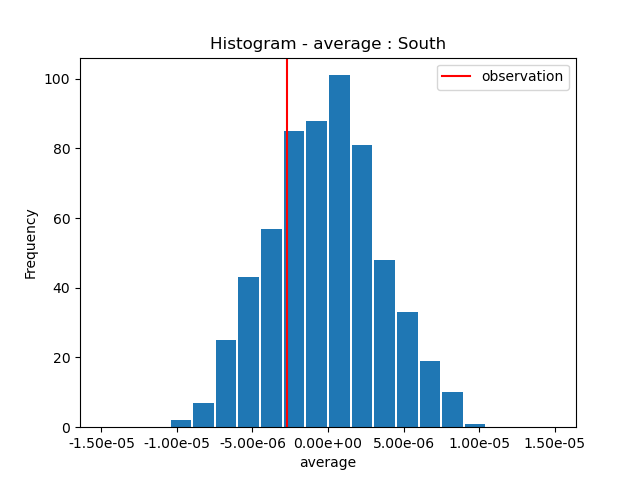}}
	\subfloat[][]{\includegraphics[width=0.25\textwidth]{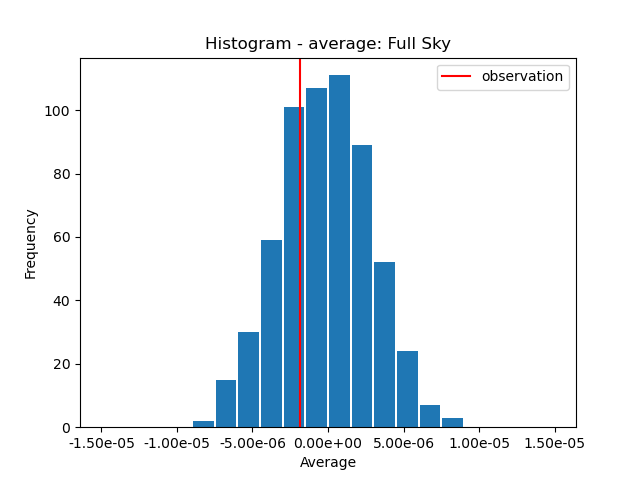}}\\
	
	\caption{Histogram of variance for the north, south and full sky, with the appropriate regions 
		masked. The histograms are computed for the map at $N = 512$.}
	\label{fig:histogram}
\end{figure*}

As the anomalies presented in the previous sections exhibit a 
dependence on the recipe for computing mean and variance, we examine the histograms 
of mean and variance of the hemispheres and the full sky at $\Res = 512$ in 
Figure~\ref{fig:histogram}. From the figure, we notice the variance of the observation in the 
northern hemisphere 
to be less than the variance from all the $600$  simulated maps, while the southern sky is 
consistent with the simulations. As a result, the full sky exhibits mildly anomalous characteristics 
with respect to the simulations. When examining the mean, it is evident that both the northern 
and the southern skies are consistent with the simulations. It has also been noted in the literature 
that at all scales the northern hemisphere 
exhibits anomalous variance with respect to the simulations, in contrast with the southern 
hemisphere, which exhibits no deviation \cite{planckIsotropy2013}. It may therefore be prudent 
to treat the hemispheres as arising from different models, when computing the mean and variance 
for normalization. 

\begin{figure*}[h!]
	\centering
	
	\subfloat[][]{\includegraphics[width=0.5\textwidth]{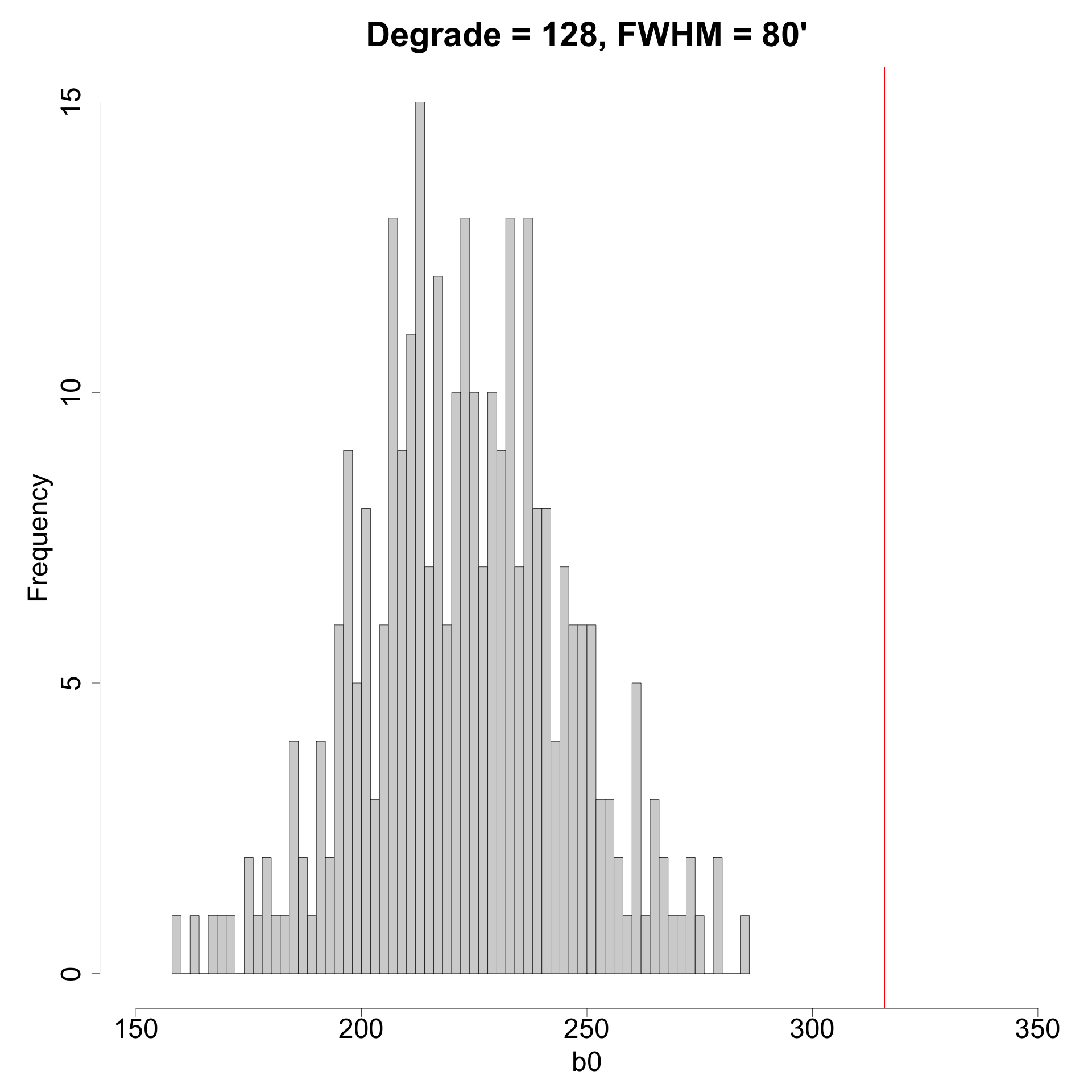}}\\
	
	\caption{Histogram of $\relBetti{0}$ values at $\nu = 0.5$ for the $\ffp$ dataset at $FWHM = 
		80'$. The distribution from the simulations indicates a tendency toward a symmetric 
		Gaussian distribution.}
	\label{fig:hist_b0_0_5}
\end{figure*}

Figure~\ref{fig:hist_b0_0_5} presents the histogram of $\relBetti{0}$ from the 
simulations for this threshold and resolution, with the observed value indicated by a red vertical 
line. It is evident from the histogram that the  distribution maybe approximated by a Gaussian 
distribution. 

\subsection{PR3 dataset}
\label{app:ffp10}
\begin{figure*}
	\centering
	
	\subfloat[][]{\includegraphics[width=0.8\textwidth]{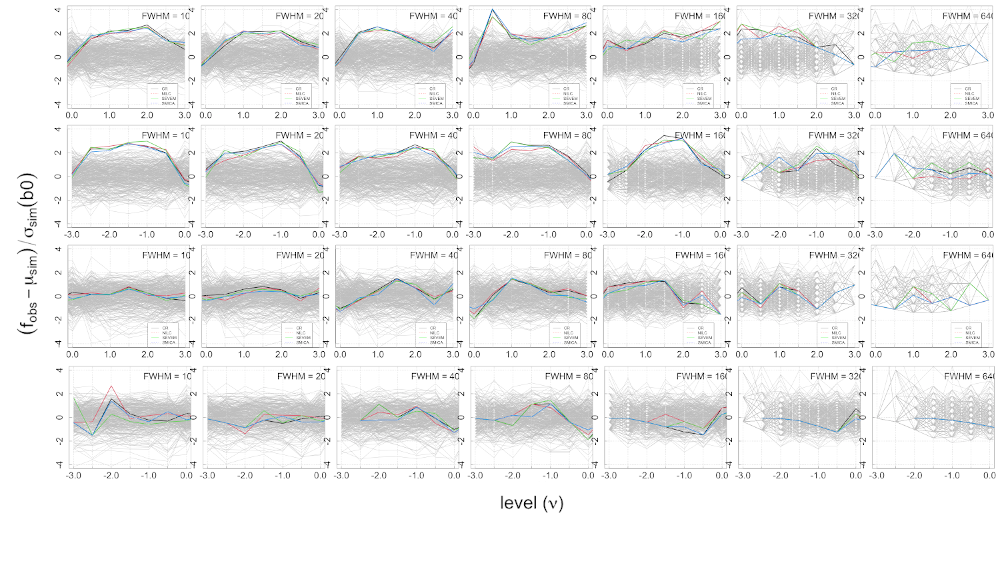}}\\
%
%
%
%
	 \caption{Graphs for local normalization. Graphs of $\relBetti{0}$ and $\relBetti{1}$  for the 
	 temperature maps for  the 
	 	PR3 dataset  for the northern (top two rows) and the southern hemisphere (bottom 
	 	two rows)The variance is computed for each hemisphere separately from the unmasked pixels 
		in that hemisphere. The graphs present the normalized differences, and each panel presents 
		the graphs for a range of degradation and smoothing scales. The mask used is the PR3 
		temperature common mask. }
	\label{fig:ffp10_betti_avgSep}
\end{figure*}

\begin{figure*}[h!]
	\centering
	
	\subfloat[][]{\includegraphics[width=0.8\textwidth]{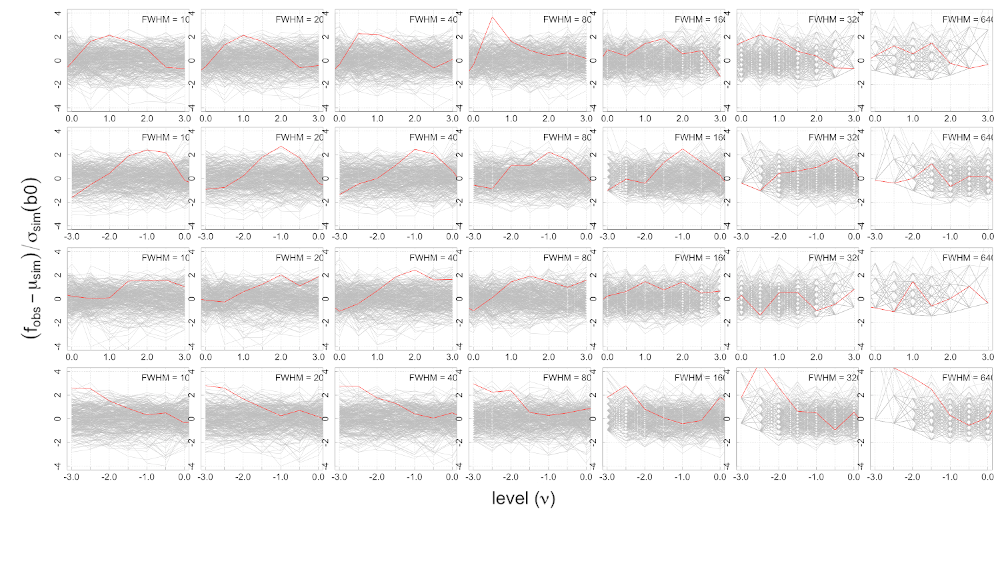}}\\
%
%
%
%
	\caption{Graphs for global normalization. Graphs of $\relBetti{0}$  and $\relBetti{1}$ for the 
	temperature maps for  the 
		PR3 dataset  for the northern (top two rows) and the southern hemispheres (bottom 
		two rows). The graphs present the normalized differences, and each panel presents the graphs 
		for a range of degradation and smoothing scales. The variance is computed from the full sky 
		from 
		the unmasked pixels. The mask used is the PR3 temperature common mask.}
	\label{fig:ffp10_betti_gAvg}
\end{figure*}

In this section, we present the graphs and table of $p$ values for the \ffp dataset. 
Figure~\ref{fig:ffp10_betti_avgSep} presents the graphs of the Betti numbers for the maps that are 
normalized with respect to local mean and variance, while Figure~\ref{fig:ffp10_betti_gAvg} 
presents the graphs for normalization where the mean and the variance as computed from the 
masked full sky. The graphs display consistent characteristics with that of the \npipe dataset 
presented in the main body of the paper, albeit with higher significance of difference between the 
data and the model. Table~\ref{tab:ffp} presents the $p$ values computed from the 
$\chi^2$-statistic. The top seven rows present the $\chi^2$-statistic computed from the combined 
level sets for a given resolution, while the bottom row presents the same for the combination of all 
thresholds at all resolutions. Commensurate with the graphs, the $p$ values for the \ffp dataset 
presents higher significance of difference between the data and the model.

\begin{table}
	\tiny
	\tabcolsep=0.09cm
	\centering
	\caption{Significance of difference for the hemispheres in PR3 dataset.}
	\subfloat[][]{\tempVarSepffp}\\
	\subfloat[][]{\tempVarGlobalffp}\\
	\tablefoot{ Two-tailed $p$ values for relative homology obtained from the 
		empirical Mahalanobis distance or $\chi^2$ test, computed from the sample covariance 
		matrices, 
		for different resolutions and smoothing scales for the PR3  dataset. Panel (a) 
		presents 
		the $p$ values for experiments where the variance is computed for each hemisphere 
		separately, and panel (b) presents results for experiments where the hemispheres are assigned 
		the variance of full sky. The last entry is the $p$ value for the summary statistic computed 
		across all resolutions. Marked in boldface are $p$ values $0.05$ or smaller.} 
	\label{tab:ffp}
\end{table}

\clearpage


\section{Results from method B}
\label{app:methodB_npipe}

In this Section, we presents thew graphs of Betti numbers and the $p$-value tables method B, 
where we extract the $\mathrm{a}_{lm}$ coefficients from the maps at the original resolution, 
convolve them with the given Gaussian beam profile, and subsequently synthesize the maps 
directly at the output resolution form the $\mathrm{a}_{lm}$'s.

\begin{figure*}[h!]
	\centering
	
	\subfloat[][]{\includegraphics[width=0.8\textwidth]{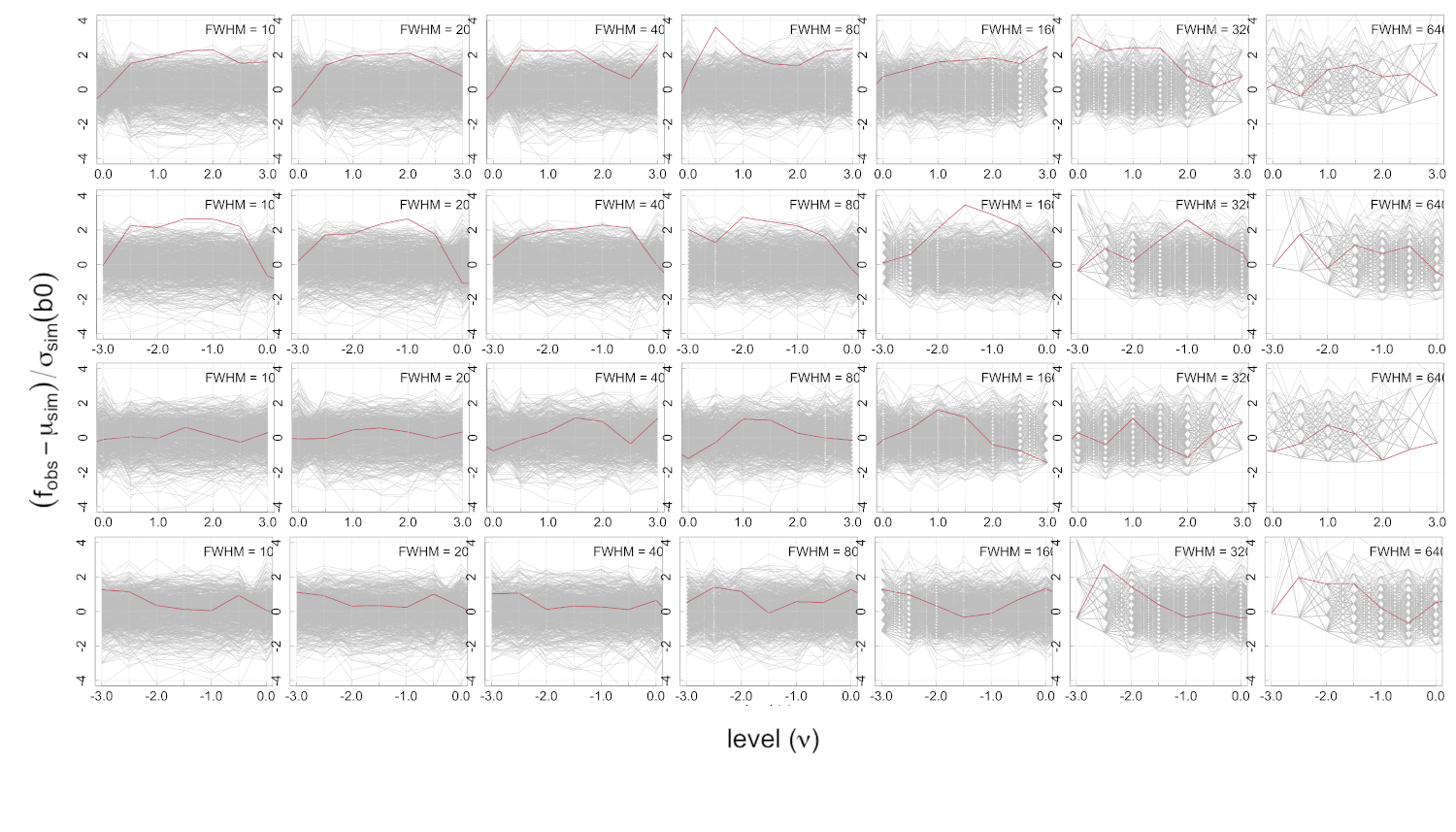}}\\
%
%
%
	\caption{Graphs of $\relBetti{0}$  and $\relBetti{1}$ for the temperature maps for  the 
		PR4 dataset  computed from method B. The top two rows present the graphs for the 
		northern hemisphere, and the bottom 
		two rows for the southern hemisphere. The graphs present the normalized differences, and 
		each panel presents the graphs 
		for a range of degradation and smoothing scales. The variance is computed from the 
		hemispheres locally from the unmasked pixels. The mask used is the PR3 temperature common 
		mask.}
	\label{fig:npipe_betti_avgSep_methodB}
\end{figure*}

\begin{table}[h!]
	\tiny
	\tabcolsep=0.09cm
	\centering
	\caption{Significance of difference for the PR4 dataset for method B.}
	\subfloat{\tempVarSepNpipeMethodB}\\
	\tablefoot{Two-tailed $p$ values for relative homology obtained from the 
		empirical Mahalanobis distance or $\chi^2$ test, computed from the sample covariance 
		matrices, 
		for different resolutions and smoothing scales for the PR4  dataset.  The last entry is the 
		$p$ value for the summary statistic computed 
		across all resolutions. Marked in boldface are $p$ values $0.05$ or smaller.} 
	\label{tab:npipe_hem_methodB}
\end{table}
\end{appendix}

\end{document}